\documentclass[twocolumn,aps,pra,groupedaddress,superscriptaddress,amsmath,amssymb,noeprint,nofootinbib,floatfix,10pt]{revtex4-2}
\usepackage[T1]{fontenc}
\usepackage{graphicx}
\graphicspath{{./Figures/}}
\usepackage{dcolumn}
\usepackage{bm}
\usepackage[dvipsnames,table,xcdraw]{xcolor}
\usepackage{hyperref}
\hypersetup{colorlinks,linkcolor={MidnightBlue},citecolor={MidnightBlue},urlcolor={MidnightBlue}}  
\usepackage{url} 
\usepackage{tabularray}
\usepackage{array}
\usepackage{makecell}

\usepackage{wrapfig}

\usepackage{tensor}

\usepackage{physics}
\usepackage{amsfonts}
\usepackage{verbatim}
\usepackage{amsmath}
\usepackage{csquotes}
\usepackage{color}
\usepackage{soul}
\usepackage{amsthm}
\usepackage{bm}
\usepackage{graphicx}
\usepackage[normalem]{ulem}
\usepackage{mathtools}

\usepackage{yfonts}

\usepackage{verbatim}

\newenvironment{equations}
{\begin{equation}\begin{aligned}}
		{\end{aligned}\end{equation}\ignorespacesafterend}

\newenvironment{equations*}
{\begin{equation*}\begin{aligned}}
		{\end{aligned}\end{equation*}\ignorespacesafterend}

\newcommand{\prt}[1]{\left(#1\right)}

\newcommand{\dg}{\dagger}
\newcommand{\tl}{\tilde}

\newcommand{\prtq}[1]{\left[#1\right]}

\newcommand{\prtg}[1]{\left\{#1\right\}}

\newcommand{\prtqB}[1]{\Bigg[#1\Bigg]}

\newcommand{\sign}{\mathrm{sign}}

\newcommand{\mcG}{\mathcal{G}}

\newcommand{\mcK}{\mathcal{K}}
\newcommand{\mcL}{\mathcal{L}}

\newcommand{\mcN}{\mathcal{N}}

\newcommand{\mcQ}{\mathcal{Q}}

\newcommand{\hA}{\hat{A}}
\newcommand{\hB}{\hat{B}}

\newcommand{\hF}{\hat{F}}

\newcommand{\hH}{\hat{H}}

\newcommand{\hV}{\hat{V}}

\newcommand{\hp}{\hat{p}}
\newcommand{\hq}{\hat{q}}

\newcommand{\hmu}{\hat{\mu}}

\newcommand{\bx}{\mathbf{x}}
\newcommand{\by}{\mathbf{y}}
\newcommand{\bz}{\mathbf{z}}

\newcommand{\bq}{\mathbf{q}}
\newcommand{\bk}{\mathbf{k}}

\newcommand{\bd}{\mathbf{d}}

\newcommand{\bv}{\mathbf{v}}
\newcommand{\bu}{\mathbf{u}}

\newcommand{\hbp}{\hat{\mathbf{p}}}
\newcommand{\hbq}{\hat{\mathbf{q}}}

\newcommand{\hbP}{\hat{\mathbf{P}}}

\newcommand{\blue}[1]{{\color{blue} #1}}

\makeatletter 

\renewcommand\onecolumngrid{
\do@columngrid{one}{\@ne}%
\def\set@footnotewidth{\onecolumngrid}
\def\footnoterule{\kern-6pt\hrule width 1.5in\kern6pt}%
}

\renewcommand\twocolumngrid{
\def\footnoterule{
\dimen@\skip\footins\divide\dimen@\thr@@
\kern-\dimen@\hrule width.5in\kern\dimen@}
\do@columngrid{mlt}{\tw@}
}%

\makeatother    

\begin{document}

\title{Gravitational Poissonian Spontaneous Localization Model of Hybrid Quantum-Classical Newtonian Gravity: Energy Increase and Experimental Bounds}

\author{Nicol\`{o} Piccione}
\email{nicolo.piccione@units.it}
\affiliation{Department of Physics, University of Trieste, Strada Costiera 11, 34151 Trieste, Italy}
\affiliation{Istituto Nazionale di Fisica Nucleare, Trieste Section, Via Valerio 2, 34127 Trieste, Italy}

\begin{abstract}
The Gravitational Poissonian Spontaneous Localization (GPSL) model is a hybrid classical–quantum framework in which Newtonian gravity emerges from stochastic collapses of a smeared mass-density operator. 
Consistency of the hybrid dynamics entails momentum diffusion and, hence, spontaneous heating.
Without smearing, which enters both the collapse (measurement) and gravitational-feedback components of the dynamics, the heating rate would be divergent.
Previous work assumed identical smearings for both components.
Here, we treat the general case of distinct spatial smearings $g_{r_C} (\bx)$ and $g_{r_G} (\bx)$, characterized, respectively, by length scales $r_C$ and $r_G$.
We characterize the spontaneous heating rate for arbitrary $g_{r_C} (\bx)$ and $g_{r_G} (\bx)$, and then discuss which smearing profiles minimize the spontaneous heating rate in relevant physical situations.
Remarkably, there are situations in which, while the measurement noise remains the same, allowing $g_{r_G} (\bx)\neq g_{r_C} (\bx)$ may reduce the feedback-induced spontaneous heating by more than 60 orders of magnitude already for $r_G = 10 r_C$.
Finally, we use our results to estimate the spontaneous heating rate of neutron stars and to set new lower bounds on the model's parameters by comparing the theoretical predictions with astronomical data on temperature, radius, and mass of neutron stars.
\end{abstract}

\maketitle

\section{Introduction\label{Sec:Introduction}}

The most common approach to merge quantum mechanics and general relativity consists in quantizing the gravitational field, producing frameworks such as String Theory and Loop Quantum Gravity, among others~\cite{Book_BambiModestoShapiroHandbookQuantumGravity2024}. However, hybrid classical-quantum theories of gravity, in which spacetime remains classical while matter is quantized, have recently emerged as serious alternatives~\cite{Tilloy2016CSLGravity,Tilloy2017LeastDecoherence,Tilloy2018GRWGravity,GaonaReyes2021GravitationalFeedback,Oppenheim2023PostQuantum,Oppenheim2023GravityTestsBounds,Piccione2025NewtonianPSL}, rather than mere semiclassical approximations. In fact, whether gravity must be quantized remains an open experimental question: proposed tests~\cite{Bose2017GravityExperiment,Marletto2017GravityExperiment,Lami2024GravityTesting,Angeli2025ProbingQuantumNatureGravity} could distinguish quantum, classical, and intermediate scenarios, but have not yet been performed.

The central theoretical challenge for hybrid approaches is consistency. The historical semiclassical gravity approach, in which classical gravity couples to the expectation value of the quantized stress-energy tensor~\cite{Book_Hu2020SemiclassicalGravity}, fails as a fundamental theory even in the Newtonian limit\footnote{In the Newtonian limit, these semiclassical models are expected to reduce to the so-called Schr\"{o}dinger-Newton equation~\cite{Bahrami2014SchrodingerNewtonEquation}.}~\cite{Eppley1977NecessityQuantumGravity,Grobardt2022MakingSenseSemiclassicalGravity}: the corresponding dynamics is nonlinear~\cite{Bahrami2014SchrodingerNewtonEquation} and allows for superluminal communication~\cite{Gisin1989StochasticDynamics}. This pathology can be avoided by introducing stochasticity. Modern hybrid classical-quantum theories rely on dynamics that can be seen as measurement-plus-feedback schemes~\cite{Kafri2014LOCCGravity,Tilloy2016CSLGravity,Tilloy2017LeastDecoherence,Tilloy2018GRWGravity,Piccione2025NewtonianPSL}, where, by construction, the reduced dynamics is linear at the density matrix level~\cite{Tilloy2024HybridDynamics,Barchielli2024HybridDynamics}. However, stochasticity entails momentum diffusion. This manifests, in general, as a spontaneous heating rate that could, in principle, be large enough to conflict with observations.

\begin{figure}
\centering
\includegraphics[width=0.45\textwidth]{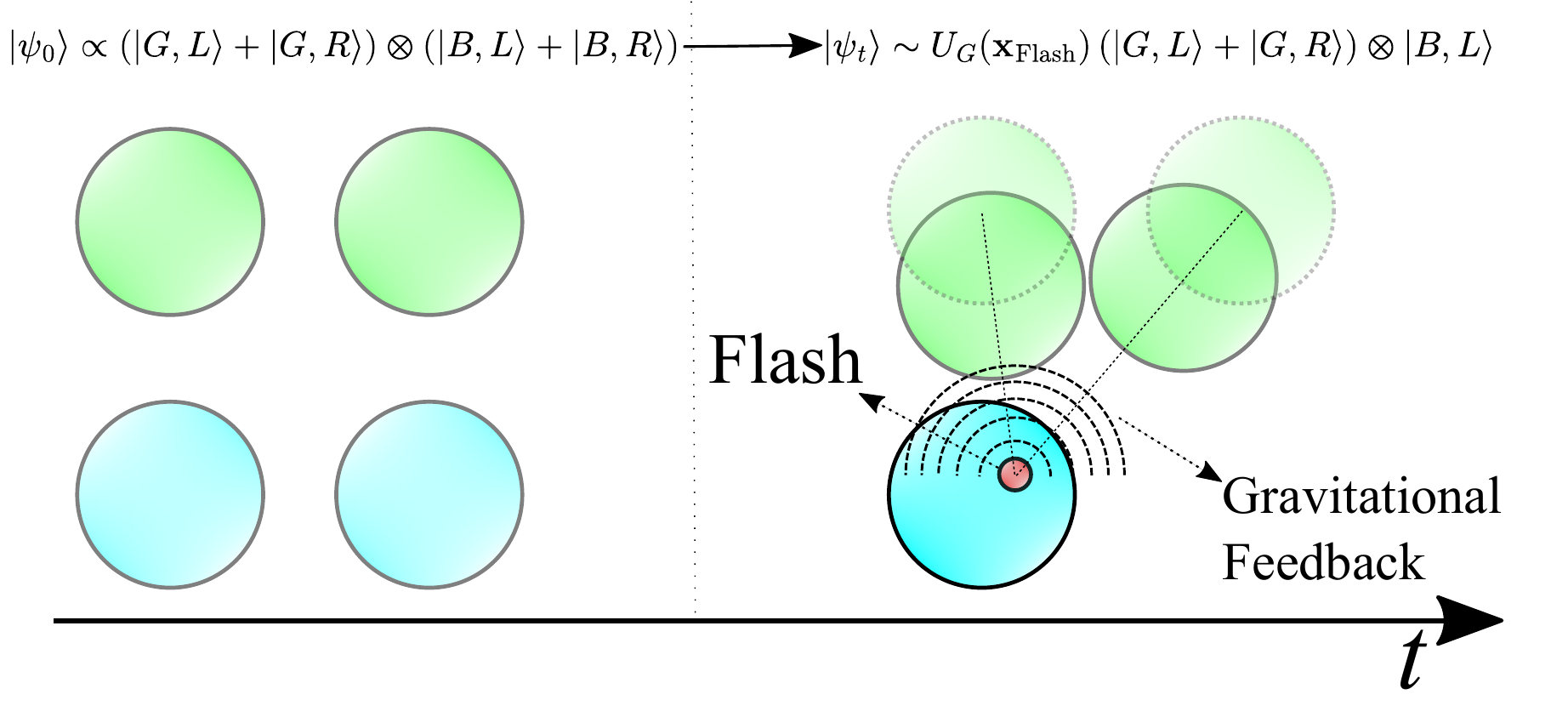}
\caption{Pictorial representation of the GPSL mechanism. Two quantum objects, green (G) and blue (B), are both in a superposition of two states: left (L) and right (R). Until there is a collapse event, the two objects do not interact gravitationally. Then, in this example, the blue object spontaneously localizes with a flash located at $\bx_{\rm Flash}$. Associated to the flash, the unitary operation $U_G (\bx_{\rm Flash})$ attracts all masses toward the collapse center.}
\label{fig:GPSLExplanation}
\end{figure}

In Ref.~\cite{Piccione2025NewtonianPSL}, we investigated the hybrid classical-quantum model of Newtonian gravity outlined in Ref.~\cite{Piccione2023Collapse} and named it Gravitational Poissonian Spontaneous Localization (GPSL). In summary, the GPSL model employs the collapse centers\footnote{Usually called \emph{flashes}, following standard terminology in the foundations of physics~\cite{Book_Tumulka2022Foundations}.} of the Poissonian Spontaneous Localization (PSL) model to source instantaneous gravitational pulses, as depicted in Fig.~\ref{fig:GPSLExplanation}. On large scales, this recovers the Newtonian phenomenology~\cite{Piccione2025NewtonianPSL}. The GPSL model has a distinctive feature compared to other approaches such as the Tilloy-Diósi model~\cite{Tilloy2016CSLGravity,Tilloy2017LeastDecoherence}: in the absence of matter, the Newtonian gravitational field vanishes \emph{exactly} rather than fluctuating around a zero average. This property may be essential for a consistent relativistic extension~\cite{Piccione2025NewtonianPSL}.

In this paper, we address a question that is critical for the viability of GPSL: whether the predicted spontaneous heating is compatible with observational data. As in other hybrid quantum-classical models of Newtonian gravity, a smearing procedure of certain operators is needed to avoid a divergent heating rate. In GPSL, the smearing is applied to the mass density operator, which is both the measured quantity and the quantity through which the gravitational feedback Hamiltonian acts. In Ref.~\cite{Piccione2025NewtonianPSL}, we assumed this smearing to be the same for both the measurement and feedback channels. Here, instead, we allow them to be different and characterized by two different smearing length scales: $r_C$ for the measurement part\footnote{The subscript \enquote{C} stands for \enquote{Collapse}.}, and $r_G$ for the gravitational feedback part\footnote{The subscript \enquote{G} stands for \enquote{Gravity}.}. 

We derive the state-dependent heating rate of GPSL for generic smearing profiles and then, following the spirit of Ref.~\cite{Piccione2026DiffusionMinimization}, we investigate which profiles, for fixed smearing lengths $r_C$ and $r_G$, minimize the heating rate. The general solution to this problem is state-dependent but we identify two classes of relevant physical situations in which we can identify state-independent answers. The first class consists of objects with dimensions much smaller than $r_C$ and $r_G$ but whose distances among each other is much higher than $r_C$ and $r_G$. 
In this case, we find that the optimal profiles are a Gaussian profile for the measurement part and a radial, increasing distribution with compact support for the gravitational feedback part. Remarkably, allowing different smearings can reduce the gravitationally-induced heating rate by more than sixty orders of magnitude already when $r_G = 10 r_C$, while the heating rate induced by the measurement part remains the same.
The second class consists of macroscopic objects much larger than $r_C$ and $r_G$, whose mass density varies on length scales much larger than $r_C$ and $r_G$, and whose particle density is high enough to contain many particles in any ball of radius $r_C$ or $r_G$ within the object.
In this case, the two optimal profiles are independent of each other. The optimal profile for the measurement part is a radial, decreasing distribution with compact support while the optimal profile for the feedback part is a constant distribution with spherical, finite support.
Cosmological considerations lead us to claim that the first class of situations is the most relevant to decide which smearing distributions should be chosen for the GPSL model.

Finally, we compare the model's predictions with observations. We estimate the GPSL-induced spontaneous heating in neutron stars and compare the theoretical predictions with astronomical data to derive both upper and lower bounds on the model's parameters, similarly to what was done in Ref.~\cite{Tilloy2019NeutronStarHeating,Adler2019NeutronStarSpontaneousHeating} for the Continuous Spontaneous Localization (CSL) and Diósi-Penrose (DP) spontaneous collapse models~\cite{Review_Bassi2003Dynamical,Review_Bassi2013Models}. The lower bounds are particularly significant: unlike theoretical lower bounds based on assumptions about collapse efficacy~\cite{Piccione2025ExploringMassDependence}, these rest directly on observational data.
In fact, as we will see, given the collapse rate $\lambda$, two terms will contribute to the spontaneous heating: one proportional to $\lambda$ and originating in the collapse mechanism and one proportional to $\lambda^{-1}$ and due to the gravitational feedback. In general, collapse models without gravitational feedback lack the latter term, so that lower bounds to $\lambda$ can only be set by requiring a certain efficacy of the collapse mechanism.
Studying astrophysical heating complements existing laboratory-based constraints on hybrid models~\cite{Oppenheim2023GravityTestsBounds,Janse2024BoundsSpacetimeDiffusion,Angeli2025ProbingQuantumNatureGravity,Angeli2025EntanglementHybridGravity} by probing a different physical regime: extreme mass densities inaccessible to terrestrial experiments.

The paper is structured as follows. In Sec.~\ref{Sec:GPSLModel}, we summarize the GPSL model focusing on its phenomenologically relevant equations. In Sec.~\ref{Sec:GPSL_HeatingRate} we derive the state-dependent heating rate formula and find the state-independent optimal distributions for minimizing the heating rate in relevant physical regimes. We also discuss why, on a cosmological scale, isolated particles give the highest contribution to the total heating rate of GPSL and why, therefore, this is the situation one should consider with the aim of minimizing the deviations induced by GPSL with respect to standard quantum mechanics. In Sec.~\ref{Sec:GPSL_NeutronStarBounds}, we estimate the heating rate for neutron stars and compare the theoretical estimations with astronomical data to put upper and lower bounds on the GPSL model parameters. Then, in Sec.~\ref{Sec:GPSL_GeneralBounds}, we merge these bounds with stronger upper bounds coming from just the measurement part~\cite{Piccione2025ExploringMassDependence} of the dynamics. Finally, we discuss the results in Sec.~\ref{Sec:Conclusions}.

\section{The GPSL Model\label{Sec:GPSLModel}}

The model investigated in Ref.~\cite{Piccione2025NewtonianPSL} is based on a previously exposed idea~\cite{Piccione2023Collapse}, which we now summarize. We first deal with a non-relativistic spontaneous collapse model~\cite{Review_Bassi2003Dynamical,Review_Bassi2013Models,Review_Bassi2023CollapseModels}, the Poissonian Spontaneous Localization (PSL) model. In this model, the probability density for a collapse to occur at spacetime point $(t,\bx)$ is proportional to the expectation value of the smeared mass density operator at that time and in that spatial location. We denote the center of such collapses by \emph{flashes}, following standard terminology in the foundations of physics~\cite{Book_Tumulka2022Foundations}. The classical gravitational field is then sourced by these flashes, that is, each of these flashes sources a gravitational pulse that acts back on the (already collapsed) matter wavefunction. Adding this gravitational feedback mechanism to the PSL model gives the hybrid classical-quantum Newtonian gravitational model that we denote by Gravitational PSL (GPSL). In this work, we only need the resulting effective dynamics, so we restrict ourselves to the minimal ingredients required later for the analysis of the heating rate; detailed derivations and further discussion can be found in Refs.~\cite{Piccione2023Collapse,Piccione2025NewtonianPSL}.

The empirical predictions of GPSL are encoded in the master equation
\begin{equation}\label{eq:GPSL_MasterEquation}
\dv{t} \rho_t = -\frac{i}{\hbar}\comm{\hH}{\rho_t} + \mcL_{\rm GPSL} [\rho_t],
\end{equation}
where
\begin{multline}
\mcL_{\rm GPSL} [\rho_t]
=
\frac{\lambda}{m_0} \int \dd[3]{\bx}\prtqB{
\\
U_G (\bx)\sqrt{\hmu_{r_C} (\bx)}\rho_t \sqrt{\hmu_{r_C} (\bx)} U_G^\dg (\bx) - \frac{1}{2}\acomm{\hmu_{r_C} (\bx)}{\rho_t}}.
\end{multline}
In the above equations, $\hH$ is the standard Hamiltonian of the system (without the gravitational part), $\hmu_{r_C} (\bx)$ is the smeared mass density operator\footnote{We denote the convolution operation by the symbol \enquote{$*$}. Therefore, $(f*\hmu)(\bx) = \int \dd[3]{\by} f(\bx-\by)\hmu(\by)$.} $\hmu_{r_C} (\bx) := (g_{r_C} * \hmu)(\bx)$, $\lambda$ is the collapse rate, $m_0$ is the proton mass, and $U_G (\bx)=\exp[-(i/\lambda \hbar) \hat{\Phi} (\bx)]$ is a unitary operator responsible for the gravitational dynamics, where $\hat{\Phi} (\bx) = - m_0 G\int \dd[3]{\by} \abs{\bx-\by}^{-1}\hmu_{r_G} (\by)$, with $G$ being the gravitational constant and $\hmu_{r_G} (\bx) := (g_{r_G} * \hmu)(\bx)$ is the smeared mass density distribution. The unitary operator $U_G (\bx)$ is what associates to each flash the gravitational impulse feedback. The PSL model can be re-obtained from the GPSL model by setting $G=0$ so that $U_G (\bx)$ becomes the identity operator. In this case, we will write $\mcL_{\rm PSL} [\rho_t]$ instead of $\mcL_{\rm GPSL} [\rho_t]$. Notice that, in contrast to Ref.~\cite{Piccione2025NewtonianPSL}, we are here accounting for the possibility that the smearing appearing inside $\hat{\Phi} (\bx)$ is different from that due to the measurement part of the dynamics. The two smearing distributions are characterized, respectively, by length scales $r_C$ and $r_G$, in the sense that their variances amount to $r_C^2$ and $r_G^2$, respectively\footnote{We denote by variance of a three-dimensional, normalized, distribution $f(\bx)$ the quantity $\sigma^2 = (1/3)\int \dd[3]{\bx} \bx^2 f(\bx)$.}.

Spontaneous collapse models and hybrid models are generally diffusive~\cite{Donadi2023DiffusiveCollapse}. In particular, in the non-relativistic regime and assuming conservation of mass, this also implies that energy is not a conserved quantity: it always increases. As we will soon see in the next section, smearing the mass density through $g_{r_C}$ and $g_{r_G}$ is necessary to avoid a divergent energy increase. The Gaussian profile has traditionally been adopted as the standard choice for smearing~\cite{Review_Bassi2003Dynamical,Review_Bassi2013Models} and, while the Gaussian possesses several convenient mathematical properties, no general physical principle is known that uniquely selects it. A first attempt at motivating the choice of smearing has been proposed in Ref.~\cite{Piccione2026DiffusionMinimization}, where the smearing profiles for the three most-studied collapse models and the Tilloy-Diósi hybrid gravitational model~\cite{Tilloy2016CSLGravity,Tilloy2017LeastDecoherence} have been selected so as to minimize the energy increase under suitable constraints. In any case, the smearings remain phenomenological ingredients of the theories, and their physical origin is presently unknown. For this reason, in what follows we keep the analysis completely general, assuming only that $g_{r_C}$ and $g_{r_G}$ are normalized, non-negative distributions characterized by the length scales $r_C$ and $r_G$, respectively.

For a system of $N$ particles, where $\hmu_{\sigma} (\bx) = \sum_j m_j g_{\sigma} (\bx-\hbq_j)$ and $M=\sum_j m_j$, the master equation becomes
\begin{multline}\label{eq:GPSL_MasterEquation_Particles}
\dv{t} \rho_t = -\frac{i}{\hbar}\comm{\hH}{\rho_t} 
-\frac{\lambda}{m_0} \prtqB{\\
	M\rho_t - \int \dd[3]{\bx} U_G (\bx)\sqrt{\hmu_{r_C} (\bx)}\rho_t \sqrt{\hmu_{r_C} (\bx)} U_G^\dg (\bx)},
\end{multline}
where
\begin{equations}
U_G (\bx) &= e^{i \sum_k r_k f_{r_G}(\hbq_k-\bx)},
\qquad
r_k:= \frac{G m_0 m_k}{\lambda \hbar},
\\
f_{r_G} (\hbq_k-\bx)
&=\int \dd[3]{\by} \frac{g_{r_G} (\hbq_k-\by)}{\abs{\bx-\by}},\\
&=\int \dd[3]{\by} \frac{g_{r_G} \prt{(\hbq_k-\bx) +\by}}{\abs{\by}}.    
\end{equations}

\section{Heating rate of the PSL and GPSL models\label{Sec:GPSL_HeatingRate}}

Both the PSL and GPSL models entail a spontaneous increase of the total energy due to the modifications with respect to standard quantum mechanics. Here, we find the relevant formulas for this heating rate. We start from the PSL case and then investigate the GPSL case. Following Ref.~\cite{Piccione2026DiffusionMinimization}, we will discuss the smearing distributions that minimize the spontaneous heating rate. Unlike models based on a weakly continuous monitoring~\cite{Piccione2026DiffusionMinimization}, however, for both PSL and GPSL the heating rate is not simply proportional to a fixed, state-independent, functional of the smearing distribution(s). Instead, it depends on the spatial distribution of the particles. Therefore, as we will show, there is no state-independent optimal profile to choose. We will identify two relevant physical regimes in which it is possible to find state-independent analytical solutions. 

\subsection{Heating rate of the PSL model\label{Subsec:PSLHeatingrate}}

Let us consider a system of $N$ particles governed by the Hamiltonian
\begin{equation}\label{eq:StandardHamiltonian}
\hH = \sum_{j=1}^N \frac{\hbp_j^2}{2 m_j} + V(\hbq_1,\dots,\hbq_N),
\end{equation}
where $m_j$ is the $j$-th particle's mass, $\hbp_j$ its momentum operator, and $\hbq_j$ its position operator. Using Eq.~\eqref{eq:GPSL_MasterEquation} with $\mcL_{\rm PSL}$ instead of $\mcL_{\rm GPSL}$ (i.e., setting $G=0$), straightforward calculations lead to the following expression for the heating rate $\dot{E}_t$:
\begin{equation}\label{eq:PSL_HeatingRate}
\dot{E}_t := \dv{t} \ev{\hH}_t = \frac{\lambda \hbar^2}{m_0} I_N [g_{r_C};\rho_t],
\end{equation}
where
\begin{equation}\label{eq:PSL_GeneralHeatingRateFunctional}
I_N [g_{r_C};\rho_t]:=
\frac{1}{2}\sum_j \frac{1}{m_j}\int \dd[3]{\bx} \Tr{\abs{\nabla_j \sqrt{\hmu_{r_C} (\bx)}}^2\rho_t},
\end{equation}
with $\nabla_j$ denoting the gradient with respect to the $j$-th particle position. Notice that the energy increase only affects the kinetic energy of the particles because $V(\hbq_1,\dots,\hbq_N)$ commutes with the additional dissipator introduced by the PSL model. If the system is composed of a single particle, the above quantity becomes state-independent:
\begin{equation}\label{eq:PSLSingleParticleHeatingFunctional}
I_1 [g_{r_C};\rho_t] 
= I[\sqrt{g_{r_C}}]
:= \frac{1}{2}\int \dd[3]{\bx} \abs{\nabla \sqrt{g_{r_C} (\bx)}}^2,
\end{equation}
where $I[f]$ is also called the Dirichlet energy of the function $f$~\cite{Book_Evans2010PartialDifferentialEquations}.
In this case, the PSL model exactly reduces to the Ghirardi-Rimini-Weber (GRW) model~\cite{Ghirardi1986Unified,Review_Bassi2003Dynamical,Review_Bassi2013Models} and the smearing distribution minimizing the spontaneous heating is the Gaussian distribution~\cite{Piccione2026DiffusionMinimization}, which gives
\begin{equation}
I[\sqrt{g_{r_C}}] 
= \frac{3}{8} r_C^{-2}
\simeq 0.375 \times r_C^{-2}.
\end{equation}

As anticipated, the heating rate of PSL is not, in general, independent of the quantum state. However, we can single out two important limit cases:
\begin{enumerate}
\item All particles are clustered within a radius much smaller than $r_C$; the system behaves as a single particle of mass $M=\sum_j m_j$~\cite{Piccione2025ExploringMassDependence}.
\item All particles are much further apart than $r_C$ from each other so that each one undergoes an independent collapse dynamics.
\end{enumerate}
Then, we get that
\begin{equations}
&\textrm{Case 1:\ }
I_N [g_{r_C};\rho_t] \rightarrow I[\sqrt{g_{r_C}}],
\\
&\textrm{Case 2:\ }
I_N [g_{r_C};\rho_t] \rightarrow N I[\sqrt{g_{r_C}}],
\end{equations}
In both these limiting cases, the minimization problem becomes the same as in the GRW model and the Gaussian smearing becomes optimal~\cite{Piccione2026DiffusionMinimization}. However, this is not true in general, as we show numerically with a two-particle counter-example in Appendix~\ref{APPSec:PSL_NonOptimality}. 
Moreover, it is quite easy to prove (see Appendix~\ref{APPSec:PSL_SandwichProof}) that $I_N [g_{r_C};\rho_t] \leq N I[\sqrt{g_{r_C}}]$. Ideally, one would also like to prove that $I_N [g_{r_C};\rho_t] \geq I[\sqrt{g_{r_C}}]$ as this would imply that the heating rate of a collection of particles is always higher than or equal to that of a single particle\footnote{While we did not prove it, we also did not manage to find a (numerical) counter-example to the inequality $I_N [g_{r_C};\rho_t] \geq I[\sqrt{g_{r_C}}]$ using a radial, decreasing, non-negative distribution of variance $r_C^2$. We conjecture that the inequality is true under these (arguably natural) assumptions.}. However, we managed to prove (see Appendix~\ref{APPSec:PSL_SandwichProof}) that [cf. Eq.~\eqref{eq:PSL_GeneralHeatingRateFunctional}]
\begin{equations}\label{eq:PSL_Inequalities}
I_N [g_{r_C};\rho_t] 
&\geq
I_{\rm CoM} [g_{r_C}; \rho_t],
\\
I[\sqrt{g_{r_C}}] 
&\geq 
I_{\rm CoM} [g_{r_C}; \rho_t],  
\end{equations}
where (with $M:= \sum_k m_k$ and $\hbP = \sum_k \hbp_k$)
\begin{equations}
I_{\rm CoM} [g_{r_C}; \rho_t] 
:&= \frac{m_0}{\lambda \hbar^2} \Tr{\frac{\hbP^2}{2M}\mcL_{\rm PSL}[\rho_t]},
\\
&= \frac{1}{2 M}\int \dd[3]{\bx} \Tr \prtg{\abs{\nabla\sqrt{\hmu_{r_C} (\bx)}}^2 \rho_t},
\end{equations}
so that $(\lambda \hbar^2/m_0) I_{\rm CoM} [g_{r_C}; \rho_t]$ gives the kinetic energy increase of the Center of Mass (CoM) of the system due to the spontaneous localization mechanism. Then, the first inequality of Eq.~\eqref{eq:PSL_Inequalities} tells us that the total heating rate is always higher than that of the CoM. Indeed, $I_N [g_{r_C},\rho_t]$ also takes into account the internal heating. The second inequality instead shows that the CoM heating is attenuated by the system containing more than one particle over lengths larger than $r_C$. In fact, $I [\sqrt{g_{r_C}}]$ is the isolated single particle heating rate and $I_{\rm CoM} [g_{r_C}; \rho_t]$ reduces to it when all particles are clustered within a spatial region characterized by a length much smaller than $r_C$. When considering a macroscopic body, $I_{\rm CoM} [g_{r_C}; \rho_t]$ shows that the CoM diffusion depends on the variation of its mass density. This qualitatively agrees with the approximate formula found in Ref.~\cite{Piccione2025ExploringMassDependence}\footnote{See Eq.~(29) of Ref.~\cite{Piccione2025ExploringMassDependence} with $\alpha=1/2$ and $\gamma_\alpha = \lambda$.} for a rigid body of constant density:
\begin{equation}
\dv{t} \ev{\frac{\hbP^2}{2M}} \propto \frac{\lambda}{r_C} \frac{A}{V},
\end{equation}
where $A$ is the surface area of the rigid body and $V$ is its volume.

Another relevant situation consists of considering an object in which each particle is (on average) localized on a scale much smaller than $r_C$, the macroscopic density $\mu_M (\bx)$ of the object varies on a scale much larger than $r_C$, and the particle density within the object is high enough that in any ball of radius $r_C$ there are many particles. In this case, we can estimate the heating rate as follows. We first rewrite Eq.~\eqref{eq:PSL_GeneralHeatingRateFunctional}:
\begin{equation}\label{eq:MacroObjectPSLContribution}
I_N [g_{r_C};\rho_t]
=
\frac{1}{2}\sum_j m_j \int \dd[3]{\bx} \Tr{\frac{\abs{\nabla g_{r_C} (\hbq_j-\bx)}^2}{4\hmu_{r_C} (\bx)}\rho_t}.
\end{equation}
Since we assume that every particle is localized on a length scale much smaller than $r_C$, we can make the substitution $g_{r_C} (\hbq_j-\bx) \rightarrow g_{r_C} (\ev{\hbq_j}-\bx)$ where $\ev{\hbq_j}$ is the average position of the $j$-th particle. Then, using the assumption that, for any $\bx$, many particles appreciably contribute to the evaluation of $\ev{\hmu_{r_C} (\bx)}$, we can make the substitution $\ev{1/\hmu_{r_C} (\bx)} \simeq 1/\ev{\hmu_{r_C} (\bx)}$ because we can neglect the fluctuations in $\hmu_{r_C} (\bx)$. Finally, for all $\bx$ in the bulk (away from the surface), we can make the substitution $1/\ev{\hmu_{r_C} (\bx)} \simeq 1/\mu_M (\bx)$. To sum up, the above quantity can be estimated by making the substitution
\begin{equation}
\Tr{\frac{\abs{\nabla g_{r_C} (\hbq_j-\bx)}^2}{4\hmu_{r_C} (\bx)}\rho_t}
\rightarrow
\frac{\abs{\nabla g_{r_C} (\ev{\hbq_j}-\bx)}^2}{4 \mu_M (\bx)}.
\end{equation}
In doing so, we identified $\ev{\hmu_{r_C} (\bx)}$ with $\mu_M (\bx)$, which only works well in the inside of the macroscopic object but does not work near (on the $r_C$ length scale) the surface of it. Moreover, while $\ev{\hmu_{r_C} (\bx)}$ is never zero, $\mu_M (\bx)$ is zero outside of the macroscopic boundaries of the macroscopic object. Therefore, the above approximation should be used only when $\bx \in V_M$, where $V_M$ denotes the spatial volume occupied by the macroscopic object.
Then, one can substitute the summation in Eq.~\eqref{eq:MacroObjectPSLContribution} with an integration over the classical mass density within the object's volume $V_M$ and, exploiting again the fact that $\mu_M (\bx)$ varies over scales much larger than $r_C$ while $\abs{\nabla g_{r_C} (\ev{\hbq_j}-\bx)}^2$ decays over this length scale, one gets 
\begin{multline}\label{eq:PSL_MacroscopicBodyHeatingRate}
I_N [g_{r_C};\rho_t]
\simeq
\frac{1}{2}\sum_j m_j \int_{V_M} \dd[3]{\bx} \frac{\abs{\nabla g_{r_C} (\ev{\hbq_j}-\bx)}^2}{4 \mu_M (\bx)}
\simeq
\\
\simeq
\frac{1}{2} \int_{V_M} \dd[3]{\by} \mu_M (\by) \int_{V_M} \dd[3]{\bx} \frac{\abs{\nabla g_{r_C} (\by-\bx)}^2}{4 \mu_M (\bx)}
\simeq
\\
\simeq
\frac{1}{8} \int_{V_M} \dd[3]{\by} \frac{\mu_M (\by)}{\mu_M (\by)} \int_{V_M} \dd[3]{\bx} \abs{\nabla g_{r_C} (\by-\bx)}^2
\simeq\\
\simeq
\frac{V_M}{8} \int \dd[3]{\bx} \abs{\nabla g_{r_C} (\bx)}^2
=
V_M I_{r_C},
\end{multline}
where we defined the state independent quantity $I_{r_C}:= (1/8) \int \dd[3]{\bx} \abs{\nabla g_{r_C} (\bx)}^2$. To extract $\mu_M (\bx)$ from the second integral, we assumed that it varies over lengths much higher than $r_C$, which cannot be true at the boundary of the macroscopic body. Thus, there should be boundary corrections to the above formula that takes into account this fact. However, the percentage of macroscopic volume over which the approximations we made do not work is roughly given by $r_C A/V_M$, where $A$ is the surface area of the macroscopic body. Indeed, for a truly macroscopic body, this will be much smaller than one. 

There are some important facts to notice about the formula $I_N [g_{r_C};\rho_t]\simeq V_M I_{r_C}$ which we are now going to discuss:
\begin{itemize}
\item The heating rate is independent of the mass density profile of the object under consideration.
\item The heating rate is proportional to the volume $V_M$ of the macroscopic object, meaning that\footnote{Neglecting losses due to thermal radiation and other energy decreasing channels.} the temperature increase rate of objects of constant density is actually independent of their volume or shape. In fact, denoting by $T$ an object's temperature, by $C(T)$ its heat capacity, and by $s(T)=C(T)/V_M$ its volumetric heat capacity, one has that
\begin{equation}
\dv{t} T 
= \frac{\dot{E}_t}{C(T)}
\simeq \frac{\lambda \hbar^2}{m_0} \frac{V_M I_{r_C}}{s(T) V_M}
= \frac{\lambda \hbar^2}{m_0}\frac{I_{r_C}}{s(T)}.
\end{equation}
\item The heating rate is again proportional to a state-independent functional of $g_{r_C} (\bx)$. Thus, a state-independent optimal profile for $g_{r_C} (\bx)$ can be found. However, in this case $\dot{E}_t \propto I[g_{r_C}]$ instead of $\dot{E}_t \propto I[\sqrt{g_{r_C}}]$ [as it was in Eq.~\eqref{eq:PSLSingleParticleHeatingFunctional}]. In this case, the optimization problem is equivalent to that solved for the CSL model in Ref.~\cite{Piccione2026DiffusionMinimization}, with solution
\begin{equation}\label{eq:PSLMacroObjectMinimizer}
g_{r_C} (\bx) = \frac{105}{32\pi (3 r_C)^7}[(3 r_C)^2 - \bx^2]^2_+,
\end{equation}
where $[x]_+ := \max \prtg{0,x}$. Indeed, computing $I_{r_C}$ with $g_{r_C} (\bx)$ being a Gaussian or that of Eq.~\eqref{eq:PSLMacroObjectMinimizer} we get:
\begin{equations}\label{eq:PSL_MacroscopicBodyOptimalHeatingRateFunctional}
\qquad
\text{Gaussian:}\
I_{r_C} &= \frac{1}{8} \frac{3 r_C^{-5}}{16 \pi^{3/2} }
\simeq
(4.2 \times 10^{-3}) r_C^{-5},
\\
\qquad
\text{Eq~\eqref{eq:PSLMacroObjectMinimizer}:}\
I_{r_C} &= \frac{1}{8} \frac{35 r_C^{-5}}{486 \pi}
\simeq
(2.87 \times 10^{-3}) r_C^{-5},
\end{equations}
confirming what found in Ref.~\cite{Piccione2026DiffusionMinimization}: taking a Gaussian in place of Eq.~\eqref{eq:PSLMacroObjectMinimizer} increases $I_{r_C}$ by $47\%$. 
\item Finally, we observe that the collective behavior of the localization mechanism\footnote{This collective mechanism is necessary to preserve the symmetrization and anti-symmetrization properties of bosons and fermions.} greatly reduces the heating rate. In fact, denoting by $n_M$ the particle density of a macroscopic object with $N_M$ particles, assuming $n_M$ to be constant (so that $N_M=n_M V_M$), and choosing $g_{r_C}$ to be Gaussian, we have that
\begin{equation}\label{eq:PSL_HeatingRateAttenuationMacroscopicObject}
\frac{N I[\sqrt{g_{r_C}}]}{V_M I_{r_C}}    
=
16 \pi^{3/2} n_M r_C^3 
\gg 1,
\end{equation}
because $n_M r_C^3 \gg 1$ by assumption.
\end{itemize}

\subsection{Heating rate of the GPSL model}

Quantifying the energy increase due to the GPSL model is not as straightforward as in the PSL case. The main problem comes from the jump operators entirely encoding the gravitational dynamics~[cf. Eq.~\eqref{eq:GPSL_MasterEquation_Particles}]. Thus, it is not at first glance obvious how to separate the actual energy increase from the variations of the gravitational potential energy or any equivalent concept for the GPSL model. In Appendix~\ref{APPSec:GPSL_EnergyIncrease}, we have investigated how the expectation value of the standard Hamiltonian (without gravity) increases with time, finding that
\begin{equation}\label{eq:GPSL_TotalEnergyIncrease}
\dot{E}_t 
= \frac{\lambda \hbar^2}{m_0} I_N [g_{r_C};\rho_t] 
+\mcG_\lambda [g_{r_C},g_{r_G};\rho_t]
-\dv{t} \ev{\hV_{r_C,r_G}}_t.
\end{equation}
The first term is exactly the PSL contribution [cf. Eq.~\eqref{eq:PSL_HeatingRate}].
The second term in Eq.~\eqref{eq:GPSL_TotalEnergyIncrease} is given by
\begin{equation}\label{eq:GPSL_GravitationalContributionHeatingRate}
\mcG_{\lambda}= \frac{G^2 m_0}{\lambda}\sum_{j,k}m_j m_k I^{(G)}_{j,k},    
\end{equation}
where 
\begin{equation}\label{eq:GPSL_HeatingFunctional}
I^{(G)}_{j,k}:= \frac{1}{2}\int \dd[3]{\bx} \Tr \{ g_{r_C} (\hbq_j-\hbq_k+\bx) \abs{\nabla f_{r_G} (\bx)}^2 \rho_t \},
\end{equation}
and $f_{r_G} (\bx) =\int \dd[3]{\by} \abs{\bx-\by}^{-1} g_{r_G} (\by)$. Indeed, $\mcG_{\lambda} \geq 0$, even for a single particle.
This second term stems from the additional noise due to the gravitational feedback.
Finally, the last term corresponds to the variation of the expectation value of the operator:
\begin{equation}\label{eq:GPSL_EffectiveGravitationalPotential}
\hV_{r_C,r_G} 
:= \int \dd[3]{\bx}\dd[3]{\by} \prtq{\frac{-G}{\abs{\bx-\by}}} \frac{\hmu_{r_C} (\bx)\hmu_{r_G}(\by)}{2}.
\end{equation}
This operator corresponds to the gravitational potential energy associated with the smeared mass density operators $\hmu_{r_C}$ and $\hmu_{r_G}$.
We argue that the expectation value of this quantity should be considered as the equivalent of the average, \emph{physical}, gravitational potential energy and, therefore, its variation rate should be ignored when searching for the optimal smearing distributions\footnote{Indeed, $\dv{t} \ev{\hV_{r_C,r_G}}_t$ can be positive, so it would even be mathematically unclear how to pose an optimization problem.}. In other words, the physical average energy is given by $E^{\rm Phys.}_t:= \langle \hH+ \hV_{r_C,r_G}\rangle_t$ where $\hH$ is given in Eq.~\eqref{eq:StandardHamiltonian}.

We now argue more in detail why one should think of $\langle\hV_{r_C,r_G}\rangle$ as the expectation value of the \emph{physical} gravitational potential energy.
First, let us consider a system composed of $N$ particles and whose dynamics are generated by the Hamiltonian
\begin{equation}
\hH_{r_C,r_G} = \sum_{j=1}^N \frac{\hbp_j^2}{2 m_j} + V(\hbq_1,\dots,\hbq_N) + \hV_{r_C,r_G},
\end{equation}
that is, the Hamiltonian of Eq.~\eqref{eq:StandardHamiltonian} plus $\hV_{r_C,r_G}$.
The dynamics would be unitary and the total energy would be conserved.
Moreover, at large distances (large with respect to $r_C$ and $r_G$), the dynamics would be close to that of a standard quantum mechanics with the addition of the quantum Newtonian potential. In fact, for $N$ particles, if it is possible to only consider large distances, it follows that\footnote{The reason why we can ignore the terms with $j=k$ is that they give a contribution proportional to the identity operator. This is explained more in detail in Ref.~\cite{Piccione2025NewtonianPSL}.}
\begin{multline}
\hV_{r_C,r_G}
=\\
\sum_{j\neq k} m_j m_k \int \dd[3]{\bx}\dd[3]{\by} \prtq{\frac{-G}{\abs{\bx-\by}}} \frac{g_{r_C} (\bx-\hbq_j) g_{r_G}(\by-\hbq_k)}{2}
\\
\approx
-\frac{G}{2} \sum_{j \neq k} \frac{m_j m_k}{\abs{\hbq_j-\hbq_k}},
\end{multline}
i.e., $\hV_{r_C,r_G}$ becomes the quantum Newtonian pair potential. We also notice that the operator $\hV_{r_C,r_G}$ is symmetric under particles exchange and, under the assumption that $g_{r_C}$ and $g_{r_G}$ are isotropic smearings, it is also translationally and rotationally invariant.
Finally, the instantaneous change in the expectation value of $\hV_{r_C,r_G}$ generated by $\hH_{r_C,r_G}$ is exactly the same as the quantity $\dv{t}\langle\hV_{r_C,r_G}\rangle_t$ in Eq.~\eqref{eq:GPSL_TotalEnergyIncrease}, given the same state $\rho_t$ at time $t$. So, the instantaneous variation $\dv{t}\langle\hV_{r_C,r_G}\rangle_t$ can be entirely attributed to a modified (on and below the lengthscales $r_C$ and $r_G$) but still unitary quantum Newtonian dynamics.

The heating rate we want to minimize is given by
\begin{equation}\label{eq:GPSL_PhysicalHeatingRate}
\dot{E}^{\rm Phys.}_t =
\frac{\lambda \hbar^2}{m_0} I_N [g_{r_C};\rho_t] 
+ \mcG_\lambda [g_{r_C},g_{r_G};\rho_t].
\end{equation}
Since it depends on the actual state of the system, it is useful to analyze two limit cases as we did for the PSL model: (1) All particles are much closer to each other than $r_C$ and $r_G$, and (2) all particles are at a much higher distance from each other than both $r_C$ and $r_G$. Then, we have\footnote{Hereafter we drop the superscript \enquote{Phys.} to lighten the notation.}
\begin{equations}\label{eq:GPSL_LimitCasesEnergyIncrease}
&\textrm{1)\ }
\dot{E}_t = \frac{\lambda \hbar^2}{m_0} I[\sqrt{g_{r_C}}] + \frac{G^2 m_0}{\lambda}M^2 I_0^{(G)}[g_{r_C},g_{r_G}],
\\
&\textrm{2)\ }
\dot{E}_t = N\frac{\lambda \hbar^2}{m_0} I[\sqrt{g_{r_C}}] + \frac{G^2 m_0}{\lambda}\tl{M}^2 I_0^{(G)}[g_{r_C},g_{r_G}],
\end{equations}
where $M=\sum_k m_k$, $\tl{M}:=\sqrt{\sum_k m_k^2}$, and 
\begin{equation}\label{eq:GPSL_GravFeedbackFunctional}
I_0^{(G)}[g_{r_C},g_{r_G}] 
:= 
\frac{1}{2} \int \dd[3]{\bx} g_{r_C} (\bx) \abs{\nabla f_{r_G} (\bx)}^2.
\end{equation}
The first equation of \eqref{eq:GPSL_LimitCasesEnergyIncrease} comes from considering $\hbq_j-\hbq_k \sim 0$ in Eq.~\eqref{eq:GPSL_HeatingFunctional} while the second one comes from neglecting $I^{(G)}_{j,k}$ when $j\neq k$.

\begin{figure}
\centering
\includegraphics[width=0.45\textwidth]{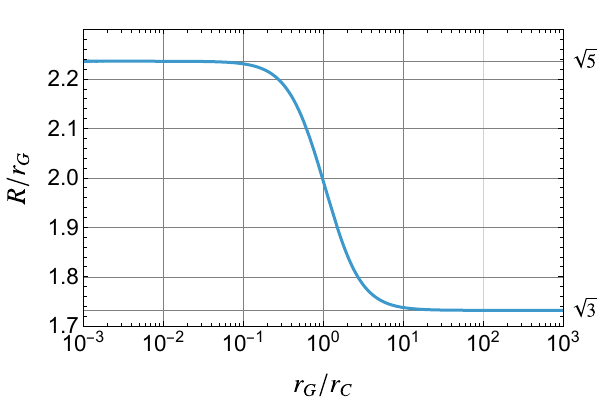}
\caption{The plot shows the ratio $R/r_G$ obtained by solving Eq.~\eqref{eq:GPSL_DifferentialEquationBoundRadius}.
The two horizontal lines represent the constant values $\sqrt{5}$ and $\sqrt{3}$.}
\label{fig:RadiusBoundOptimalGPSL}
\end{figure}

\begin{figure}
\centering
\includegraphics[width=0.45\textwidth]{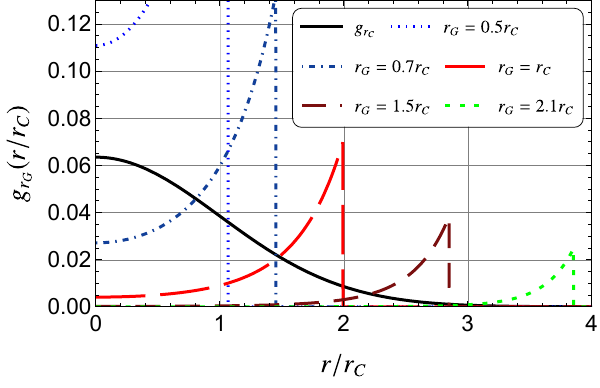}
\caption{This plot compares $g_{r_G}$ [Eq.~\eqref{eq:GPSL_OptimalDistribution}] to $g_{r_C}$ (black continuous line, a Gaussian distribution with variance $r_C^2$), for different values of $r_G$. Here, $r=\abs{\bx}$.}
\label{fig:SmearingFunctionsPlot}
\end{figure}

Since $M\rightarrow 0$ or $\tl{M}\rightarrow 0$ would reduce the above optimization problem to the PSL one, we deem reasonable to fix $g_{r_C}$ as the Gaussian distribution with variance $r_C^2$. The problem then becomes that of minimizing $I_0^{(G)}$ when $g_{r_C} (\bx)$ is a Gaussian of variance $r_C^2$ and we fix the variance $r_G^2$ of the to-be-found smearing distribution $g_{r_G} (\bx)$. In Appendix~\ref{APPSec:GPSL_Optimization}, we show that in this case the optimal distribution $g_{r_G} (\bx)$ is 
\begin{equation}\label{eq:GPSL_OptimalDistribution}
g_{r_G} (\bx) = \frac{e^{-R^2/2r_C^2}}{R^3} \frac{\bx^2+3 r_C^2}{4\pi r_C^2}e^{\bx^2/2 r_C^2}\Theta (R-\abs{\bx}),
\end{equation}
where $R$ is a function of $r_G$ and $r_C$, numerically determined by solving for $y$ the equation
\begin{equation}\label{eq:GPSL_DifferentialEquationBoundRadius}
2y-2+\frac{3}{y}-\frac{3\sqrt{\pi}}{2}\frac{e^{-y}}{y^{3/2}} \textrm{erfi} \prt{\sqrt{y}}
\!=\! 
3 \prt{\frac{r_G}{r_C}}^2\!\!\!,
\quad\!\!
R=r_C \sqrt{2y}.
\end{equation}
The resulting ratio $R/r_G$ is shown in Fig.~\ref{fig:RadiusBoundOptimalGPSL}.
Fig.~\ref{fig:SmearingFunctionsPlot} shows a plot of $g_{r_G}$ compared to $g_{r_C}$ for various values of $r_G/r_C$.
More generally, the optimal $g_{r_G} (\bx)$ can be found as a function of any arbitrary $g_{r_C} (\bx)$ (see Appendix~\ref{APPSec:GPSL_Optimization}). The general solution is
\begin{equation}
g_{r_G} (r)
=
\frac{g_{r_C} (R)}{4\pi R^3 r^2}\dv{r}\prt{\frac{r^3}{g_{r_C} (r)}}\Theta(R-r),
\end{equation}
where $r=\abs{\bx}$ and $R$ is found by solving the equation
\begin{equation}
R^2 - 2\frac{g_{r_C} (R)}{R^3}\int_0^{R} \frac{r^4}{g_{r_C} (r)} = 3 r_G^2.
\end{equation}

\begin{figure}
\centering
\includegraphics[width=0.48\textwidth]{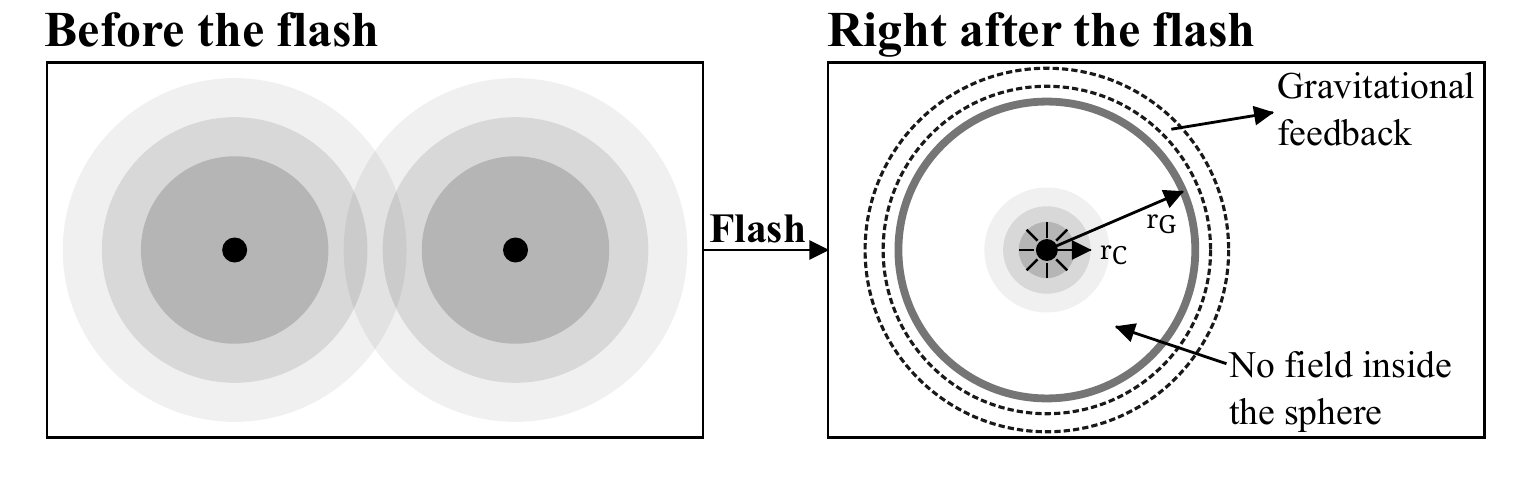}
\caption{Pictorial representation of the GPSL mechanism for a single particle and the $g_{r_G}$ of Eq.~\eqref{eq:GPSL_OptimalDistribution}. Before the flash, the particle is in a generic quantum state. In the picture, a superposition of states with spread much larger than $r_C$. When the flash occurs, the particle localizes around the flash on a length $r_C$. The gravitational feedback field is generated by a mass density with shape $g_{r_G}$ centered on the flash. Using the $g_{r_G}$ of Eq.~\eqref{eq:GPSL_OptimalDistribution} (see also Fig.~\ref{fig:SmearingFunctionsPlot} for $r_G=2.1 r_C$), there is basically no back-reaction on the spontaneously localized particle.}
\label{fig:GPSLExplanationFeedbackOptimization}
\end{figure}

The fact that the optimal $g_{r_G} (\bx)$ [Eq.~\eqref{eq:GPSL_OptimalDistribution}] is an increasing function with compact support may seem strange at first but it is actually physically reasonable. As explained in Ref.~\cite{Piccione2023Collapse} and depicted in Fig.~\ref{fig:GPSLExplanationFeedbackOptimization}, each flash acts as a (smeared) source of the gravitational field for an infinitesimal time.
Mathematically, we are effectively optimizing for a single collapsing particle and, after each collapse, the particle is mostly contained within a length comparable with $r_C$ of the flash. As the particle does not feel its own standard Newtonian field, we can identify the entire gravitational field generated by the flash with the gravitational back-reaction noise. 
Importantly, the gravitational field generated by a spherical shell of uniform mass density is zero in the space enclosed by the shell. Therefore, under the variance and normalization constraints, the intuition would be to concentrate as much mass as possible away from the flash. Putting all the mass on a thin shell would give the distribution:
\begin{equation}
g_R (r) = \frac{\delta (r-R)}{4\pi R^2},
\qquad
R = \sqrt{3} r_G,
\end{equation}
which explains the limit $r_G \gg r_C$ of Fig.~\ref{fig:RadiusBoundOptimalGPSL}. However, one can improve on this intuitive result by allowing $R$ to be larger and distributing the mass over a non perfectly thin shell. In particular, the functional $I_0^{(G)}[g_{r_C},g_{r_G}]$ of Eq.~\eqref{eq:GPSL_GravFeedbackFunctional} penalizes the gravitational field generated by the feedback profile in the region where the post-collapse particle is likely to be found, namely the region weighted by $g_{r_C}$. Since the gravitational field at radius $r$ is controlled by the feedback-generating-mass (determined by $g_{r_G}$) enclosed inside $r$, the minimization favors profiles for which this feedback-generating-mass remains small in the high-probability region of the localized particle and then rises rapidly near the outer radius allowed by the variance constraint. This is why the optimal profile is radially increasing and with compact support.

\begin{figure}
\centering
\includegraphics[width=0.45\textwidth]{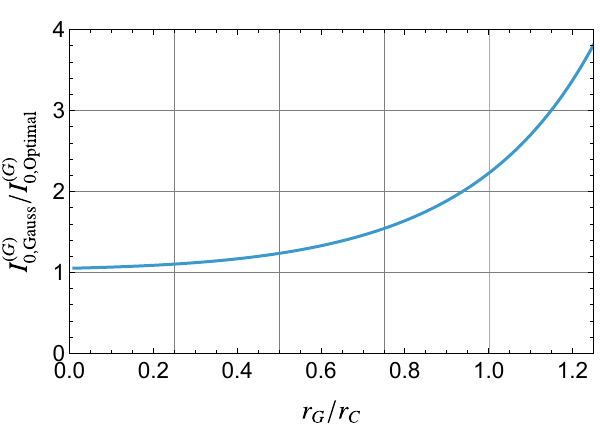}
\caption{The plot shows the ratio of $I_0^{(G)}$ computed with Gaussian distributions over $I_0^{(G)}$ computed with Gaussian smearing for the measurement part and the optimal distribution for the gravitational feedback. We set $r_C=1$ to make the numerical calculations. This plot confirms that this ratio is always greater than one.}
\label{fig:RatioGaussOverOptimal}
\end{figure}

Numerically [see Fig.~\ref{fig:RadiusBoundOptimalGPSL}], we see that $R(r_G,r_C)\sim 2 r_G$, with $R(r_G,r_C) \simeq \sqrt{5} r_G$ for $r_G \ll r_C$ and $R(r_G,r_C) \simeq \sqrt{3} r_G$ for $r_G \gg r_C$ (see Fig.~\ref{fig:RadiusBoundOptimalGPSL}). Using the Gaussian smearing in place of that of Eq.~\eqref{eq:GPSL_OptimalDistribution} leads to an increase of the heating rate that depends on the ratio $r_G/r_C$ (see Fig.~\ref{fig:RatioGaussOverOptimal}). In fact, computing $I_0^{(G)}$ with both distributions being Gaussian gives [see Appendix~\ref{APPSec:GPSL_Optimization}]
\begin{equation}
I_0^{(G)} 
= 
\frac{1}{2 r_C^4}\prtq{1+\frac{2}{\pi}\prt{\frac{\eta^2}{\sqrt{1+2\eta^2}}-2 \arctan\sqrt{1+2\eta^2}}},
\end{equation}
where $\eta:= r_C/r_G$. Using, instead, the optimal distribution of Eq.~\eqref{eq:GPSL_OptimalDistribution} gives\footnote{A longer analytical and closed expression for $I_0^{(G)}$ can be found in Eq.~\eqref{APPeq:GPSL_OptimalHeatingFunctionalClosedExpression} of Appendix~\ref{APPSec:HeatingFunctionalCalculation}.}
\begin{multline}
I_0^{(G)}
=
\frac{1}{r_C^4 \sqrt{2\pi}} 
\times\prtqB{\\
	\frac{e^{-2y}}{(2y)^3} \int_0^{\sqrt{2y}} \dd{u} u^4 e^{u^2/2} + \int_{\sqrt{2y}}^{\infty} \dd{u} \frac{e^{-u^2/2}}{u^2}}.
\end{multline}
When $r_G=r_C$, the ratio between $I_0^{(G)}$ computed with a Gaussian $g_{r_G}$ and the optimal $g_{r_G}$ quantifies to approximately $2.235$. Fig.~\ref{fig:RatioGaussOverOptimal} shows how this ratio seems to be monotonically increasing and very fast for $r_G > r_C$. For example, when $r_G = 10 r_C$, the ratio between $I_0^{(G)}$ computed with the Gaussian and with the optimal distribution gives roughly $2.57 \times 10^{62}$, which means that using the Gaussian distribution hugely overestimates the actual heating required by the (generalized) model. As for PSL, the minimization that holds for a single particle or for the limit cases of Eq.~\eqref{eq:GPSL_LimitCasesEnergyIncrease} already fails for two particles, as we show in Appendix~\ref{APPSec:GPSL_CounterExample}.

\begin{figure}
\centering
\includegraphics[width=0.45\textwidth]{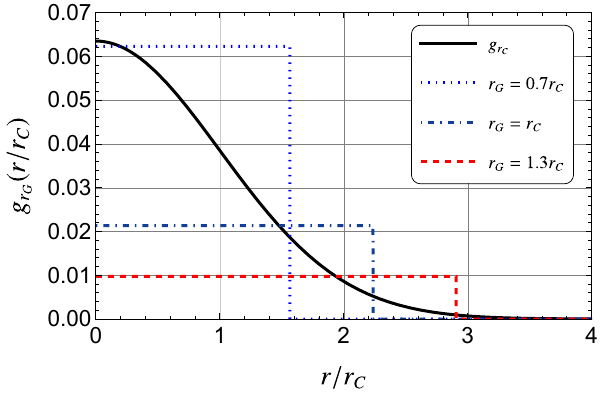}
\caption{This plot compares $g_{r_G}$ [Eq.~\eqref{eq:GPSL_OptimalMacroscopicBodySmearingFeedback}] to $g_{r_C}$ (black continuous line, a Gaussian distribution of variance $r_C^2$), for different values of $r_G$. Here, $r=\abs{\bx}$.}
\label{fig:SmearingFunctionsPlotLargeBody}
\end{figure}

The other relevant situation that we analyzed in Sec.~\ref{Subsec:PSLHeatingrate} for the PSL heating rate is that of a macroscopic object in which each particle is (on average) localized on a scale much smaller than $r_C$, the macroscopic density $\mu_M (\bx)$ of the object varies on a scale much larger than $r_C$, and the particle density within the object is high enough that in any ball of radius $r_C$ there are many particles. Assuming that all this also applies compared to $r_G$, we can derive an approximate formula for the heating rate of such an object. To do so, we start by approximating Eq.~\eqref{eq:GPSL_HeatingFunctional} as
\begin{equation}
I^{(G)}_{j,k}
\simeq
\frac{1}{2}\int \dd[3]{\bx} g_{r_C} (\ev{\hbq_j}-\ev{\hbq_k}+\bx) \abs{\nabla f_{r_G} (\bx)}^2,
\end{equation}
where we are exploiting the fact that every particle is very well localized on the $r_C$ length scale. So, $I^{(G)}_{j,k}$ only depends on the distance between the average position of the two particles. Hereafter, for notational convenience, we will write $I^{(G)} (\bd)$ instead of $I^{(G)}_{j,k}$, where
\begin{equation}
I^{(G)} (\bd)
=
\frac{1}{2}\int \dd[3]{\bz} g_{r_C} (\bd+\bz) \abs{\nabla f_{r_G} (\bz)}^2,
\end{equation}
and $\bd = \ev{\hbq_j}-\ev{\hbq_k}$.
The gravitational contribution to the heating rate can then be estimated as follows [cf. Eq.~\eqref{eq:GPSL_GravitationalContributionHeatingRate}]:
\begin{multline}\label{eq:GPSL_NeutronStarGravitationalContribution}
\mcG_{\lambda} 
= 
\frac{G^2 m_0}{\lambda} \sum_{j} m_j  \sum_k m_k I^{(G)}_{j,k}
\sim
\\
\sim
\frac{G^2 m_0}{\lambda} \sum_{j} m_j  \int \dd[3]{\by} \mu_M (\by) I^{(G)} \prt{\by-\ev{\hbq_j}}
\sim
\\
\sim
\frac{G^2 m_0}{\lambda} \int \dd[3]{\bx} \dd[3]{\by} \mu_M (\bx) \mu_M (\by) I^{(G)} \prt{\bx-\by}.
\end{multline}
We expect the function $I^{(G)} (d)$ to rapidly decay over lengths between $r_C$ and $r_G$. Over such lengths, it is perfectly reasonable to assume that the mass density of the macroscopic object does not vary appreciably, thus allowing further approximations:
\begin{equation}\label{eq:GPSL_MacroscopicBodyGravitationalContributionApproximated}
\mcG_{\lambda} 
\sim
\frac{G^2 m_0}{\lambda} I_{r_G} \times \int \dd[3]{\bx} \mu_M^2 (\bx),
\end{equation}
where we defined $I_{r_G}:= \int \dd[3]{\by} I^{(G)} \prt{\by}$. This quantity can be simplified (see Appendix~\ref{APPSec:GPSL_MacroscopicBodyHeatingRate}) to
\begin{equation}
I_{r_G}
=
2\pi \int_0^{\infty} \dd{r} \frac{G^{2}_{r_G} (r)}{r^2},
\end{equation}
where $G_{r_G} (r):= 4\pi \int_0^{r} \dd{s} s^2 g_{r_G} (s)$, which is the radial cumulative distribution associated to the smearing distribution $g_{r_G} (\bx)$. Remarkably, the quantity $I_{r_G}$ does not depend on $g_{r_C}$ at all but only on the gravitational smearing distribution $g_{r_G}$. Finding the distribution $g_{r_G} (r)$ that minimizes $I_{r_G}$ leads to the same problem already solved in Appendix~\ref{APPSec:GPSL_Optimization} but with $g_{r_C} (\bx) = 1$ (see Appendix~\ref{APPSec:GPSL_MacroscopicBodyHeatingRate}). The optimal distribution is
\begin{equation}\label{eq:GPSL_OptimalMacroscopicBodySmearingFeedback}
g_{r_G} (\bx) = \prt{\frac{4\pi}{3}R^3}^{-1}\Theta\prt{R-\abs{\bx}},
\qquad
R=\sqrt{5}r_G,
\end{equation}
that is, a constant distribution with spherical finite support (see Fig.~\ref{fig:SmearingFunctionsPlotLargeBody}). Using this optimal distribution, one gets
\begin{equation}\label{eq:GPSL_OptimalMacroscopicBodyHeatingFunctional}
I_{r_G}
=
\frac{12 \pi}{5 \sqrt{5}} r_G^{-1}
\simeq
3.37 \times r_G^{-1},
\end{equation}
while using a Gaussian distribution gives
\begin{equation}
I_{r_G}
=
2 \sqrt{\pi} r_G^{-1}
\simeq
3.54 \times r_G^{-1}.
\end{equation}
Therefore, allowing a different smearing for the gravitational feedback does not appreciably decrease the heating rate due to the gravitational feedback.

\subsection{Choosing the smearing distribution(s)\label{Subsec:CosmologicalConsiderationsHeatingRateSmearingChoice}}

We have seen how different physical situations lead to different optimal spatial profiles for the smearing distributions. Here, we argue that, if the goal is to minimize violation of (local) energy conservation across the universe, the most important regime to consider is that of individual particles that are much further apart than both $r_C$ and $r_G$. Therefore, should one make a choice about the smearing distributions, these should be the Gaussian distribution for the measurement part and by Eq.~\eqref{eq:GPSL_OptimalDistribution} for the gravitational feedback part. On the other hand, if one makes the reasonable assumption that the two smearing distributions should be equal, then one should choose the Gaussian smearing, as the measurement part gives most of the heating rate for the kinds of particles that make up most of the mass in the Universe, as we argue through heuristic estimations in the following.

First, because gravitational experiment have confirmed Newton's law down to $\sim 50$ µm~\cite{Tan2020ImprovementInverseSquareLaw,Lee2020NewTestInverseSquareLaw}, we consider that the maximum value of both $r_C$ and $r_G$ is $10^{-4}$ meters. Since energy non-conservation is arguably the most prominent phenomenological consequence of hybrid models, if smearing distributions have to be chosen, one should consider which physical regime is the most relevant for cosmological considerations. This is, by far, that of isolated particles. In fact, the mean baryon number density $n_b$ in the Universe is $n_b \simeq 0.25 \rm{m}^{-3}$~\cite{Review_PDG2024}\footnote{In particular, see~\cite{PDG2025AstrophysicalConstantsParameters}.}. This means that the average inter-particle spacing $d$ can be estimated as $d \sim n_b^{-1/3} \sim 1.6 \rm{m}$, which is much larger than $10^{-4}$ meters. Moreover, the contribution of matter in objects with high particle density is attenuated by the collective localization mechanism [cf. Eq.~\eqref{eq:PSL_HeatingRateAttenuationMacroscopicObject}].

Starting from Eq.~\eqref{eq:GPSL_PhysicalHeatingRate}, one can estimate the ratio of the measurement and gravitational feedback contribution to the heating rate in the case of $g_{r_C}=g_{r_G}$ and an isolated particle of mass $m$ as
\begin{equation}\label{eq:HeatingRateContributionsRatio}
\frac{\dot{E}_t^{\rm PSL}}{\mcG_\lambda}
\sim
\lambda^2 r_C^2 \frac{\hbar^2}{m_0^2 G^2 m^2},
\end{equation}
where we used the fact that $I[\sqrt{g_{r_C}}] \sim r_C^{-2}$ and $I_0^{(G)} \sim r_C^{-4}$. Looking at Fig.~\ref{fig:exclusionPlotGeneral} in Sec.~\ref{Sec:GPSL_GeneralBounds}, we see that the lowest value of the product $\lambda^2 r_C^2$ is attained for $\lambda \sim 10^{-18} \rm{Hz}$ and $r_C \sim 10^{-4} \rm{m}$. Using these values and inserting the proton mass $m=m_0$, we get that
\begin{equation}
\frac{\dot{E}_t^{\rm PSL}}{\mcG_\lambda}
\sim
3.19 \times 10^{15},
\end{equation}
thus proving that, in the cosmological context and considering baryonic matter, the highest contribution to the heating rate comes from the localization mechanism and not from the gravitational feedback noise.

It is currently believed that most of the matter in the Universe is made of so-called cold dark matter~\cite{Review_PDG2024}\footnote{In particular, see~\cite{PDG2025DarkMatter}.}. Since GPSL is a gravitational model, it makes sense to assume that it affects dark matter. Even in this case, the by far most relevant physical situation is that of isolated particles on the $10^{-4}$ meter length scale. In fact, denoting by $\rho_{\rm DM}$ the mass density of dark matter and by $m_{\rm DM}$ its mass, one has that the interparticle spacing $d_{\rm DM}$ can be estimated as
\begin{equation}
d_{\rm DM}
\sim 
\prt{\frac{m_{\rm DM}}{\rho_{\rm DM}}}^{1/3}.
\end{equation}
Demanding that $d_{\rm DM} \gg 10^{-4} \rm{m}$ means demanding that $m_{\rm DM} \gg 10^{-12} \rho_{\rm DM} \rm{m}^{3}$. In Ref.~\cite{PDG2025AstrophysicalConstantsParameters}, we see that the cosmological density of dark matter is $\rho_{\rm  DM} \simeq 2.3 \times 10^{-27} \rm{Kg}\ \rm{m}^{-3}$, which means that we should demand
\begin{equation}
m_{\rm DM}
\gg
2.3 \times 10^{-39} \rm{Kg}
\simeq
1.2 \times 10^{-3} \rm{eV}/c^2.
\end{equation}
For fermionic dark matter, the lower bound on the particle mass is $70 \rm{eV}$~\cite{PDG2025DarkMatter}, which is of course much higher than $1.2 \times 10^{-3} \rm{eV}$. For bosonic dark matter, instead, the lower bound is $10^{-22} \rm{eV}$, which would allow for non-isolated particles on the $r_C$ and $r_G$ scale. As far as we know, there are no general upper bounds to the mass of a dark matter particle. However, it seems reasonable to assume that also for dark matter the most important contribution to the heating rate would come from the localization mechanism as they should have a mass more than $10^{7}$ times that of a proton to make the quantity of Eq.~\eqref{eq:HeatingRateContributionsRatio} close enough to one. Alternatively, they should be clumped into micro-objects with such high masses, which seems improbable given that dark matter is supposed to (practically) only interact gravitationally.

Let us stress that the above cosmological estimates should be understood as heuristic motivation for considering the isolated-particle regime as the most relevant one. A fully systematic treatment is, in fact, outside the scope of this paper. Taking the above analysis for good, cosmological considerations motivate using the isolated-particle regime as the benchmark for selecting the smearing profiles that minimize the energy increase, i.e., a Gaussian $g_{r_C}$ for the measurement channel and the isolated-particle optimum for $g_{r_G}$ [Eq.~\eqref{eq:GPSL_OptimalDistribution}] for the gravitational feedback channel, or a Gaussian if one enforces equal smearings.

\section{Neutron star spontaneous heating in the GPSL model\label{Sec:GPSL_NeutronStarBounds}}

In Ref.~\cite{Tilloy2019NeutronStarHeating}, astronomical data from neutron stars have been used to constrain the Continuous Spontaneous Localization (CSL) and Diósi-Penrose (DP) models of spontaneous wavefunction collapse. Eqs.~\eqref{eq:GPSL_LimitCasesEnergyIncrease} and~\eqref{eq:GPSL_MacroscopicBodyGravitationalContributionApproximated} show that the heating rate due to the gravitational feedback may become extremely high when considering objects with an extreme mass density such as neutron stars. In this section, we follow the same kind of analysis made in Ref.~\cite{Tilloy2019NeutronStarHeating} to obtain an estimation of observational bounds on the GPSL model.

\begin{table}[t]
\centering
\resizebox{0.45\textwidth}{!}{%
	\begin{tabular}{|l|l|l|}
		\hline
		& PSR J1840-1419         & PSR J2144–3933          \\ \hline
		Radius          & $10$ km                & $13$ km                 \\ \hline
		Mass            & $1 M_\odot$            & $1.4 M_\odot$           \\ \hline
		Temperature     & $0.28 \times 10^{6}$ K & $4.2 \times 10^{4}$ K   \\ \hline
		Radiation Power & $4.38\times 10^{23}$ W & $3.75 \times 10^{20}$ W \\ \hline
	\end{tabular}%
}
\caption{Astronomical data and estimations for neutron stars PSR J1840-1419~\cite{Tilloy2019NeutronStarHeating} and PSR J2144–3933~\cite{Guillot2019ColdestNeutronStar}. $M_\odot$ ($\simeq 1.988 \times 10^{30}\ \rm{kg}$) denotes a solar mass. For the radius of PSR J2144–3933~\cite{Guillot2019ColdestNeutronStar}, we took the largest allowed value as it leads to more conservative bounds [see Eq.~\eqref{eq:UpperLowerBoundsNeutronStars} for $\lambda_{-}$]. \blue{In any case,} using $10$ km in place of $13$ km does not change Fig.~\ref{fig:ExclusionPlotNeutronStars} appreciably.
\blue{Similarly, we point out that $4.2 \times 10^{4}$ K is an upper bound on the temperature of PSR J2144–3933~\cite{Guillot2019ColdestNeutronStar}.}
}
\label{tab:NeutronStarsData}
\end{table}

We take the same neutron star PSR J1840-1419~\cite{Keane2013NeutronStarJ1840} considered in Ref.~\cite{Tilloy2019NeutronStarHeating} as well as PSR J2144–3933, which (as far as we know) is the coldest neutron star observed until now~\cite{Guillot2019ColdestNeutronStar}. We consider these neutron stars as spherical objects characterized by a radius $L$, a total mass $M_N$, a temperature $T$, and a (classical) mass density distribution $\mu_N (\bx)$.  
A rough estimate of the heating rate for a neutron star can be obtained by applying the formulas of Eqs.~\eqref{eq:PSL_MacroscopicBodyHeatingRate} and~\eqref{eq:GPSL_MacroscopicBodyGravitationalContributionApproximated}. Neutron stars are not rigid bodies with atomic and molecular structures typical of matter in the solid state. Therefore, to motivate the validity of Eqs.~\eqref{eq:PSL_MacroscopicBodyHeatingRate} and~\eqref{eq:GPSL_MacroscopicBodyGravitationalContributionApproximated}, and given the values in Table~\ref{tab:NeutronStarsData}, we estimate the thermal De Broglie wavelength\footnote{Following Ref.~\cite{Rivieccio2025ThermalIndexNeutronStars}, we define the thermal De Broglie wavelength $\lambda_{\rm th}$ as $\lambda_{\rm th}:= \sqrt{(2\pi \hbar^2)/(m_N k_B T)}$, where $m_N$ is the neutron mass. Notice that, as in Ref.~\cite{Rivieccio2025ThermalIndexNeutronStars}, we use the thermal De Broglie wavelength for massive particles and not the one used for ultra-relativistic or massless particles.} $\lambda_{\rm th}$ of neutrons inside the stars to be
\begin{equations}
\text{PSR J1840-1419:}\quad 
&\lambda_{\rm th} \simeq 3.29 \times 10^{-12} \rm{m},
\\
\text{PSR J2144–3933:}\quad 
&\lambda_{\rm th} \simeq 8.48 \times 10^{-12} \rm{m}.
\end{equations}
In both cases, $\lambda_{\rm th}$ is much smaller than all considered values of $r_C$ and $r_G$ so that, on average, each particle is localized enough to justify the approximations leading to Eqs.~\eqref{eq:PSL_MacroscopicBodyHeatingRate} and~\eqref{eq:GPSL_MacroscopicBodyGravitationalContributionApproximated}. The other two assumptions required, that $\mu_N (\bx)$ varies over scales much larger than $r_C$ and $r_G$, and that the particle density is high enough are obviously met by neutron stars.

Using the values of $I_{r_C}$ and $I_{r_G}$ resulting from the optimal smearing distributions [see Eq.~\eqref{eq:PSL_MacroscopicBodyOptimalHeatingRateFunctional} and Eq.~\eqref{eq:GPSL_OptimalMacroscopicBodyHeatingFunctional}], the estimated heating rate for a neutron star is
\begin{equation}
\dot{E}_t 
\sim 
0.012 \times \frac{\lambda \hbar^2}{m_0} \frac{L^3}{r_C^5}
+
3.37 \times \frac{G^2 m_0}{\lambda r_G}\int \dd[3]{\bx} \mu_N^2 (\bx).
\end{equation}
We have now to choose a mass density profile $\mu_N (\bx)$ to employ for setting the bounds. The most conservative bound (i.e., the lowest heating rate) is obtained by using a uniform density, that is $\mu_N (\bx) = \overline{\mu}_N := M_N/V_L$, where $V_L$ is the volume of a sphere of radius $L$. This is so because
\begin{equation}\label{eq:MinimalDensityProfile}
\int_{V_L} \dd[3]{\bx} \mu_N (\bx) = M_N
\implies
\int_{V_L} \dd[3]{\bx} \mu_N^2 (\bx)
\geq
\frac{M_N^2}{V_L}.
\end{equation}
In this case, one obtains
\begin{equation}
\dot{E}_t 
\sim 
0.012 \times \frac{\lambda \hbar^2}{m_0} \frac{L^3}{r_C^5}
+
0.80 \times \frac{G^2 m_0 M_N^2}{\lambda L^3 r_G}.
\end{equation}
Using other possible distributions should not, however, qualitatively change the result. For example, using the common Tolman VII solution density profile~\cite{Raghoonundun2015TolmanSolution} for the mass density, given by\footnote{See Eq.~(4) of Ref.~\cite{Raghoonundun2015TolmanSolution}  with $\mu=1$. It shows the formula in a different but equivalent form.}
\begin{equation}
\mu_N (\bx) = \frac{5}{2} \overline {\mu} \prt{1-\frac{\bx^2}{L^2}},
\quad
\overline {\mu}:= \frac{1}{V_L}\int \dd[3]{\bx} \mu_N (\bx), 
\end{equation}
one has that the ratio between the gravitational contribution to the heating rate computed with the Tolman solution density profile or the uniform density is approximately $1.43$, i.e., it is of order unity.

\begin{figure}
\centering
\includegraphics[width=0.45\textwidth]{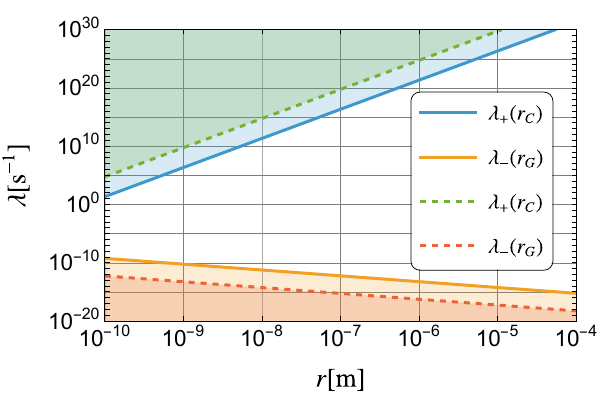}
\caption{Exclusion plot obtained using the data for neutron stars PSR J1840-1419~\cite{Tilloy2019NeutronStarHeating} (dashed lines), PSR J2144–3933~\cite{Guillot2019ColdestNeutronStar} (thick lines), and Eq.~\eqref{eq:UpperLowerBoundsNeutronStars}. The strongest upper and lower bounds come from PSR J2144–3933. Values of $r_C$ and $r_G$ higher than $10^{-4}$ m have not been considered as Newton’s inverse-square law has been directly tested at roughly gravitational strength down to $\sim 50$ µm~\cite{Tan2020ImprovementInverseSquareLaw,Lee2020NewTestInverseSquareLaw}.}
\label{fig:ExclusionPlotNeutronStars}
\end{figure}

As explained in Ref.~\cite{Tilloy2019NeutronStarHeating}, the emitted power of the neutron star can be estimated through its temperature by means of the Stefan-Boltzmann law
\begin{equation}
P_{\rm rad} = 4\pi L^2 \sigma T^4,
\qquad
\sigma = 5.6 \times 10^{-8} \rm{W}\rm{m}^{-2}\rm{K}^{-4}
\end{equation}
Assuming the star to be at thermal equilibrium, its emitted power cannot be lower than the spontaneous heating rate, thus providing bounds on the GPSL model: we must discard values of $\lambda$, $r_C$, and $r_G$ such that $\dot{E}_t \geq P_{\rm rad}$. Once $r_C$ and $r_G$ are fixed, these bounds can be easily found by solving the quadratic inequality
\begin{equation}
\prtq{0.012 \times \frac{\hbar^2}{m_0}\frac{L^3}{r_C^5}}\lambda^2 - P_{\rm rad} \lambda + \prtq{0.80\times \frac{G^2 m_0 M_N^2}{L^3 r_G}} \leq 0.
\end{equation}
However, for most of the values of $r_C$ and $r_G$ considered, the upper bound is almost entirely determined by the PSL contribution and the lower one by the gravitational feedback one. So we get
\begin{equation}\label{eq:UpperLowerBoundsNeutronStars}
\lambda_+ \simeq 1047.2 \times \frac{m_0 \sigma T^4}{\hbar^2 L} r_C^5,
\quad
\lambda_- \simeq 0.064 \times \frac{G^2 m_0 M_N^2}{\sigma T^4 L^5} r_G^{-1}.
\end{equation}
Using Eq.~\eqref{eq:UpperLowerBoundsNeutronStars} we can use the astronomical data and estimations reported in Table~\ref{tab:NeutronStarsData} to generate the bounds shown in Fig.~\ref{fig:ExclusionPlotNeutronStars}. Notice how the upper and lower bounds depend on $r_C$ and $r_G$ separately, so that they can be both represented at once in Fig.~\ref{fig:ExclusionPlotNeutronStars}.

The astronomical data used to put the bounds in Fig.~\ref{fig:ExclusionPlotNeutronStars} suffer from uncertainties and the derivation of the bounds itself involves various simplifications and heuristic estimations. 
Firstly, we are applying a non-relativistic theory (GPSL) to very relativistic objects (Neutron Stars).
Secondly, we are treating the neutron stars as perfect spherical blackbody emitters uniquely characterized by their mass, radius, and temperature.
Thirdly, we are (as commonly done in the collapse model literature~\cite{Carlesso2022Present}) performing a quite conservative estimation of the constraints due to the fact that, under the thermal equilibrium assumption, we are excluding values of $\lambda_\pm$ on the idea that if the GPSL-induced heating was too strong, the neutron star would get more energy than the radiated one, thus not being at thermal equilibrium. Assuming that, for example, at most $1\%$ of the emitted power can be due to GPSL-induced heating, the bounds would improve. In particular, $\lambda_{+} \propto P_{\rm rad}$ and $\lambda_{-} \propto P_{\rm rad}^{-1}$ so that both bounds would improve by two orders of magnitude.
Despite all of these simplifications, we stress that the goal of this section is to obtain a conservative estimation of the constraints that observational data put on the GPSL model.

As shown in Sec.~\ref{Sec:GPSL_GeneralBounds}, the upper bound $\lambda_{+}$ of Eq.~\eqref{eq:UpperLowerBoundsNeutronStars} is vastly higher (more or less eighteen orders of magnitude according to Fig.~\ref{fig:exclusionPlotGeneral}) than the upper bound coming from spontaneous radiation constraints to the PSL model~\cite{Piccione2025ExploringMassDependence}. Thus we can focus on discussing the effects of uncertainties and simplifications on $\lambda_{-}$. In particular, we can focus on PSR J2144-3933 as it gives much stronger lower bounds than PSR J1840-1419. Indeed, to obtain the most conservative bounds, we have used the upper bound temperature on PSR J2144-3933 and the highest radius. Taking lower values for either of them would increase $\lambda_{-}$ [see Eq.~\eqref{eq:UpperLowerBoundsNeutronStars}]. Analogously, if we used a different density profile than the constant one, the quantity $\lambda_{-}$ would increase [see Eq.~\eqref{eq:MinimalDensityProfile}]. For the neutron star mass, we used the value assumed in Ref.~\cite{Guillot2019ColdestNeutronStar}. Changing it to a smaller one, like $1 M_\odot$ instead of $1.4 M_\odot$, would decrease $\lambda_{-}$ by approximately $49\%$, thus leaving it in the same order of magnitude, an almost invisible change on a plot like that of Figs.~\ref{fig:ExclusionPlotNeutronStars} and~\ref{fig:exclusionPlotGeneral}. Finally, one could consider the fact that the Neutron star is not exactly at thermal equilibrium and that emitted power could deviate from the Stefan-Boltzmann law by a factor $\epsilon$, in which case $\lambda_{-}$ would vary by a factor $\epsilon^{-1}$. As long as $\epsilon \sim 1$, the plots in Figs.~\ref{fig:ExclusionPlotNeutronStars} and~\ref{fig:exclusionPlotGeneral} remain qualitatively the same.

\section{Other bounds\label{Sec:GPSL_GeneralBounds}}

\begin{figure}
\centering
\includegraphics[width=0.45\textwidth]{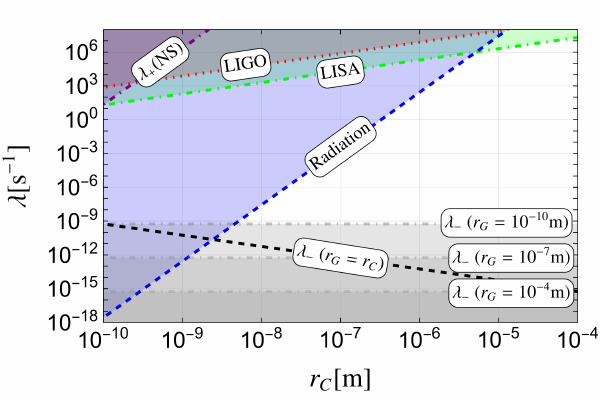}
\caption{Exclusion plot integrating the bounds of Fig.~\ref{fig:ExclusionPlotNeutronStars} with those found in Ref.~\cite{Piccione2025ExploringMassDependence}. The horizontal gray lines represent (from top to bottom) the lower bounds to $\lambda$ for $r_G = (10^{-10},10^{-7},10^{-4}) \textrm{m}$ and these are independent of $r_C$. The upper bounds are instead all functions of $r_C$: the dashed blue line corresponds to upper bounds coming from comparison with spontaneous radiation, while the dotted red line and the dot-dashed green line correspond to upper bounds coming from comparisons with experimental data of LIGO and LISA. Neglecting the black dashed line, the experimentally allowed values of $\lambda$ and $r_C$ reside within the white region lower bounded by the appropriate gray line (depending on the chosen value of $r_G$) and upper bounded by the other lines. Finally, the black dashed line represents the lower bound to $\lambda$ as a function of $r_G$ (and not of $r_C$).}
\label{fig:exclusionPlotGeneral}
\end{figure}

In Ref.~\cite{Piccione2025ExploringMassDependence}, the PSL model with Gaussian smearing is investigated as the $\alpha=1/2$ generalization of the CSL model. There, upper bounds on $\lambda (r_C)$ are derived based on experimental data from (predicted) spontaneous radiation and comparison with calibration data of the LIGO and LISA Pathfinder detectors. Theoretical lower bounds for the PSL model are also presented in the same paper, but they rest on questionable assumptions on the required efficacy of collapse models in their localization effects. Instead, our analysis of the GPSL model provides lower bounds based on observational data, thus making them more reliable for restraining a hybrid classical-quantum model of Newtonian gravity. One could also derive lower bounds based on the experimental data analyzed in Ref.~\cite{Piccione2025ExploringMassDependence}. However, we expect these lower bounds to be orders of magnitude lower than those obtained here, the reason being that these data come from earth-based experimental setups involving objects with mass densities much lower than neutron stars.

Merging the upper bounds from Ref.~\cite{Piccione2025ExploringMassDependence} with the upper and, most importantly, lower bounds coming from Sec.~\ref{Sec:GPSL_NeutronStarBounds} gives Fig.~\ref{fig:exclusionPlotGeneral}. The bounds coming from Ref.~\cite{Piccione2025ExploringMassDependence} have been obtained assuming a Gaussian smearing so that, strictly speaking, they represent an overestimation of the upper bounds on $\lambda (r_C)$. However, the bounds one would obtain, for each experimental situation there considered, by searching the smearing distribution giving the smallest deviation from standard quantum mechanics would most probably give the same qualitative picture, as shown here for the heating rate of macroscopic objects: employing Eqs.~\eqref{eq:PSL_MacroscopicBodyOptimalHeatingRateFunctional} and~\eqref{eq:GPSL_OptimalMacroscopicBodyHeatingFunctional} does not change the order of magnitude of the heating rate compared to the Gaussian smearing case.

Other experimental platforms (such as matter-wave interferometry setups, cantilever, cold atoms, etc.), can be used to constrain the parameter space of collapse models~\cite{Carlesso2022Present,Review_Carlesso2025CollapseModelsShortReview}. However, while many of these have been explored for the CSL model, they have been investigated neither here nor in Ref.~\cite{Piccione2025ExploringMassDependence} for the GPSL model.
The reason is that the physical systems studied in Ref.~\cite{Piccione2025ExploringMassDependence} already gave the strongest (upper) bounds for the CSL model and so Ref.~\cite{Piccione2025ExploringMassDependence} focused on those systems.

\section{Conclusions\label{Sec:Conclusions}}

In this work, we have analyzed the energy increase implied by the Gravitational Poissonian Spontaneous Localization (GPSL) model of hybrid classical-quantum Newtonian gravity~\cite{Piccione2023Collapse,Piccione2025NewtonianPSL}, which can be seen as the PSL spontaneous collapse model~\cite{Piccione2023Collapse,Piccione2025ExploringMassDependence} equipped with a feedback mechanism implementing the Newtonian gravitational interaction~\cite{Piccione2025NewtonianPSL}. Spontaneous heating is a generic feature of hybrid models due to the stochastic dynamics required for consistency~\cite{Galley2023ConsistentQuantumClassicalGravity}, and its magnitude determines whether such models can survive empirical scrutiny. 
Our main results are summarized in the following.

First, we derived the formula for the state-dependent spontaneous heating rate of the GPSL model with arbitrary spatial smearings $g_{r_C} (\bx)$ and $g_{r_G} (\bx)$. Finding the physical heating rate related to the gravitational feedback required the separation of the energy variation due to the non-unitary part of the dynamics into two parts: a term that represents the variation of the gravitational potential energy, and the actual heating rate.

Second, following Ref.~\cite{Piccione2026DiffusionMinimization}, we identified the state-independent spatial profiles $g_{r_C} (\bx)$ and $g_{r_G} (\bx)$ that minimize the heating rate in relevant physical regimes: isolated elementary or composite particles and macroscopic objects with high enough particle density. Interestingly, in the cosmologically relevant\footnote{See discussion in Sec.~\ref{Subsec:CosmologicalConsiderationsHeatingRateSmearingChoice}.} case of isolated particles, not setting $g_{r_G} (\bx)=g_{r_C} (\bx)$ can reduce the gravitationally-induced heating rate by more than sixty orders of magnitude already when $r_G = 10 r_C$. We recall, however, that the dominant contribution to the heating rate comes from the spontaneous collapse mechanism and not from the gravitational feedback.

Third, we estimated the spontaneous heating of neutron stars and placed upper and lower bounds by comparing the estimations with astronomical data on the star PSR J2144–3933~\cite{Guillot2019ColdestNeutronStar}, the coldest known neutron star. Then, we merged these new bounds with previous bounds on the PSL model~\cite{Piccione2025ExploringMassDependence}, thus obtaining an exclusion plot [Fig.~\ref{fig:exclusionPlotGeneral}] for the GPSL model. The allowed parameter region remains substantial, leaving GPSL viable. The lower bounds we added in this paper deserve particular emphasis since previous theoretical lower bounds on collapse models~\cite{Piccione2025ExploringMassDependence} rely on assumptions about the required efficacy of localization; assumptions that are difficult to justify rigorously. In contrast, our lower bounds follow directly from observational data.

The present work should be viewed as a further step in assessing the GPSL model as a candidate hybrid classical-quantum theory of Newtonian gravity. In Ref.~\cite{Piccione2025NewtonianPSL}, we showed that the vacuum is stable in the  GPSL model, in the sense that to the absence of mass corresponds an \emph{exactly} (i.e., non-fluctuating) vanishing gravitational field, a property that may be relevant for a relativistic extension of such models~\cite{Piccione2025NewtonianPSL}. Here, we have shown that the associated spontaneous heating is not in conflict with present observational data coming from astronomical observations.

As a final comment, let us discuss the non-relativistic character of the GPSL model. The GPSL model is intrinsically non-relativistic as it is built on the non-relativistic PSL collapse dynamics and the feedback mechanism implements Newtonian gravity. Indeed, building a complete relativistic extension is obviously very challenging as it would constitute a successful merging of gravity and quantum mechanics. An intermediate step could consist of exploring the first-order post-Newtonian corrections to the model. This has already been done for the Schr\"{o}dinger-Newton equation, e.g. in Refs.~\cite{Manfredi2015SchrodingerNewtonBeyondNewton,Brizuela2022PostNewtonianSchrNewt} or for other hybrid approaches~\cite{Pietsch2026pPostNewtonianGravitationalCollapseModel}. The same kind of extensions may be applicable to GPSL without losing its advantages over other approaches. In such a case, one should compare the predictions of a post-Newtonian GPSL with those of other hybrid approaches and those of linearized quantum gravity, for which a number of experimental signatures that cannot be obtained by the quantum Newtonian interaction have been predicted~\cite{Chen2025QuantumEffectsBeyondNewtonPotential,Lantano2026AngularMomentumGRFrameDragging}. Moreover, there may be astrophysical systems featuring strong gravitational effects that may be used to put further constraints on a possible post-Newtonian GPSL model. In fact, predictions of a post-Newtonian GPSL model could be tested against data on astrophysical systems in post-Newtonian regimes such as certain binary systems of compact objects~\cite{Review_Barack2019BlackHolesGravWavesFundPhys}.

\section*{Acknowledgements}

N.P. thanks A. Bassi for numerous and useful comments about the draft of this work.
N. P. acknowledges support from the PNRR MUR project PE0000023-NQSTI, INFN, and the University of Trieste. This project has received funding from the European Union’s Horizon Europe research and innovation programme under the Marie Skłodowska-Curie Grant Agreement No. 101150889 (CPQM).

\bibliography{Bibliography}

\clearpage
\onecolumngrid
\appendix





\section{Non-existence of a universal state-independent optimal smearing distribution for the PSL model\label{APPSec:PSL_NonOptimality}}

Here we show, by means of a counter-example, that there is no state-independent optimal distribution for minimizing the heating rate of the PSL model.

Let us consider the case in which we have two particles with masses $m_1 = m_0$ and $m_2=10 m_0$. They are in a product state and each of them is almost perfectly localized around a certain position so that they are separated by a distance $d$. By changing coordinates inside the integrals of Eq.~\eqref{eq:PSL_GeneralHeatingRateFunctional} and going to spherical coordinates, the quantity to minimize becomes
\begin{equation}
I_N [g_{r_C};\rho_t]= \frac{\pi}{4}\int_0^{\infty} \dd{r} \int_{0}^\pi \sin(\theta)r^2 [g_{r_C}' (r)]^2\prtq{
	\frac{m_1}{m_1 g_{r_C} (r) + m_2 g_{r_C} (\tl{r})}
	+
	\frac{m_2}{m_2 g_{r_C} (r) + m_1 g_{r_C} (\tl{r})}
},
\end{equation}
where $\tl{r} = \sqrt{r^2 + d^2 +2 r d \cos(\theta)}$.
Setting $r_C=d=1$ just for the numerical evaluation, the above quantity gives around $0.4136$ when inserting a Gaussian distribution. Instead, using the sub-Gaussian distribution $C_p \exp{-(r/\alpha_p r_C)^p}$ where\footnote{Here $\Gamma(x)$ is the Gamma-function, not a decoherence rate.} $\alpha_p = \sqrt{3 \Gamma(3/p)/\Gamma(5/p)}$ and $C_p = p (4\pi \Gamma(3/p) \alpha_p^3 r_C^3)^{-1}$ with $p=1.9$ gives around $0.4125$, which is clearly lower than in the Gaussian case. Since the Gaussian is optimal when $d=0$, this proves that there is no universal optimal smearing distribution.

\clearpage
\section{Proof of the PSL inequality\label{APPSec:PSL_SandwichProof}}

Here, we prove that for any normalized, positive distribution $g_{r_C} (\bx)$, one has
\begin{equation}\label{APPeq:SandwichPSL}
\text{(1):}
\quad I_N [g_{r_C};\rho_t] \leq N I[\sqrt{g_{r_C}}],
\qquad
\text{(2):}\quad
I_N [g_{r_C};\rho_t]
\geq
I_{\rm CoM} [g_{r_C}; \rho_t],
\qquad
\text{(3):}\quad
I[\sqrt{g_{r_C}}] \geq I_{\rm CoM} [g_{r_C}; \rho_t],
\end{equation}
where
\begin{equations}
I[\sqrt{g_{r_C}}] 
&= \frac{1}{2}\int \dd[3]{\bx} \abs{\nabla \sqrt{g_{r_C} (\bx)}}^2
= 2 \pi \int_0^\infty \dd{r} r^2 \frac{[g_{r_C}'(r)]^2}{4 g_{r_C} (r)},
\\
I_N [g_{r_C};\rho_t]
&=
\int_{\mathbb{R}^{3N}} \dd{y} \mcK_N [g_{r_C},y] \rho_t(y;y),
\\
I_{\rm CoM} [g_{r_C}; \rho_t]
&=
\frac{1}{2 M}\int \dd[3]{\bx} \Tr \prtg{\abs{\nabla\sqrt{\hmu_{r_C} (\bx)}}^2 \rho_t},
\end{equations}
and we introduced the quantity
\begin{equation}
\mcK_N [g_{r_C},y] 
:= 
\sum_j \mcQ_N^{(j)} [g_{r_C},y],
\qquad
\mcQ_N^{(j)} [g_{r_C},y] 
:=
\frac{1}{2 m_j} \int \dd[3]{\bx} \abs{\nabla_j \sqrt{\mu_{r_C} (\bx;y)}}^2
=
\frac{m_j}{2} \int \dd[3]{\bx}\frac{\abs{\nabla g_{r_C} (\bx-\by_j)}^2}{4 \mu_{r_C} (\bx;y)},
\end{equation}
with $\mu_{r_C} (\bx;y)=\sum_k m_k g_{r_C} (\bx-\by_k)$ and $y:= \prtg{\by_1,\dots,\by_N}$.

\subsection{Proof of inequality (1)}

Since $m_k \geq 0$ and $g_{r_C} (\bx) \geq 0$, one has that
\begin{equation}
\mcQ_N^{(j)} [g_{r_C},y] \leq \frac{m_j}{2} \int \dd[3]{\bx}\frac{\abs{\nabla g_{r_C} (\bx-\by_j)}^2}{4 m_j g_{r_C} (\bx-\by_j)}
=
I[\sqrt{g_{r_C}}].
\end{equation}
This immediately implies that
\begin{equation}
\mcK_N [g_{r_C},y] = \sum_j \mcQ_N^{(j)} [g_{r_C},y]  \leq \sum_j I[\sqrt{g_{r_C}}] = N I[\sqrt{g_{r_C}}]
\implies
I_N [g_{r_C};\rho_t] \leq N I[\sqrt{g_{r_C}}].
\end{equation}

\subsection{Proof of inequality (2)}

We can define the vector 
\begin{equation}
\bv_j (\bx;y) := \nabla_j \sqrt{\mu_{r_C} (\bx;y)},
\qquad
\mcK_N [g_{r_C},y]  := \frac{1}{2}\int \dd[3]{\bx} \sum_j \frac{\abs{\bv_j (\bx;y)}^2}{m_j}.
\end{equation}
Pointwise, the Cauchy–Schwarz inequality\footnote{The Cauchy–Schwarz inequality for real vectors says that $(\bu\cdot\bv)^2 \leq \bu^2 \bv^2$. Assume now that $u_j = a_j/\sqrt{b_j}$ and $v_j = \sqrt{b_j}$ where the $a_j$ are real numbers while the $b_j$ are non-negative real numbers. Then, the inequality becomes $\prt{\sum_j a_j}^2 \leq \prt{\sum_j a_j^2/b_j}\prt{\sum_k b_k}$, which can be inverted to write $\prt{\sum_j a_j^2/b_j} \geq \prt{\sum_k b_k}^{-1} \prt{\sum_j a_j}^2$.} gives us that
\begin{equation}
\sum_j \frac{\abs{\bv_j (\bx;y)}^2}{m_j}
\geq
\frac{1}{M}\abs{\sum_j \bv_j (\bx;y)}^2
=
\frac{1}{M}\abs{-\nabla \sqrt{\mu_{r_C} (\bx;y)}}^2,
\end{equation}
which implies that 
\begin{multline}
\mcK_N [g_{r_C},y] \geq \frac{1}{2 M} \int \dd[3]{\bx} \abs{\nabla \sqrt{\mu_{r_C} (\bx;y)}}^2,
\implies
\\
\implies
I_N [g_{r_C};\rho_t]
\geq
I_{\rm CoM} [g_{r_C}; \rho_t]
=
\frac{1}{2 M}\int_{\mathbb{R}^{3N}} \dd{y} \int \dd[3]{\bx}\abs{\nabla \sqrt{\mu_{r_C} (\bx;y)}}^2 \rho_t (y;y).
\end{multline}

\subsection{Proof of inequality (3)}

This last inequality can also be proved by means of a pointwise inequality. First, we notice that
\begin{equation}
\nabla \mu_{r_C} (\bx;y)
=
\mu_{r_C} (\bx;y) \sum_k w_k (\bx;y) \nabla \ln(g_{r_C} (\bx-\by_k)),
\qquad
w_k (\bx; y) := \frac{m_k g_{r_C} (\bx-\by_k)}{\mu_{r_C} (\bx;y)}.
\end{equation}
Notice that $w_k (\bx;y) \in [0,1]$ and $\sum_k w_k (\bx;y) = 1$.
Because of convexity of the norm, one has that
\begin{multline}
\abs{\nabla \sqrt{\mu_{r_C} (\bx;y)}}^2
=
\frac{1}{4 \mu_{r_C} (\bx;y)}\abs{\nabla \mu_{r_C} (\bx;y)}^2
\leq
\frac{\mu_{r_C} (\bx;y)}{4} \sum_k w_k (\bx;y) \abs{\nabla \ln(g_{r_C} (\bx-\by_k))}^2
=\\
=
\frac{1}{4} \sum_k m_k \frac{\abs{\nabla g_{r_C} (\bx-\by_k)}^2}{g_{r_C} (\bx-\by_k)}.
\end{multline}
Then, we have that
\begin{multline}
I_{\rm CoM} [g_{r_C}; \rho_t]
=
\frac{1}{2 M}\int_{\mathbb{R}^{3N}} \dd{y} \dd[3]{\bx}\abs{\nabla \sqrt{\mu_{r_C} (\bx;y)}}^2 \rho_t (y;y)
\leq
\frac{1}{8 M}\int_{\mathbb{R}^{3N}} \dd{y} \dd[3]{\bx} \prtq{\sum_k m_k \frac{\abs{\nabla g_{r_C} (\bx-\by_k)}^2}{g_{r_C} (\bx-\by_k)}} \rho_t (y;y)
=\\
=
\frac{1}{8 M}\int_{\mathbb{R}^{3N}} \dd{y} \sum_k m_k \prtq{\int \dd[3]{\bx}\frac{\abs{\nabla g_{r_C} (\bx)}^2}{g_{r_C} (\bx)}} \rho_t (y;y)
=
\frac{1}{8}\int_{\mathbb{R}^{3N}} \dd{y} \prtq{\int \dd[3]{\bx} \frac{\abs{\nabla g_{r_C} (\bx)}^2}{g_{r_C} (\bx)}} \rho_t (y;y)
=\\
=
\frac{1}{2} \int \dd[3]{\bx} \abs{\nabla \sqrt{g_{r_C} (\bx)}}^2
=
I[\sqrt{g_{r_C}}].
\end{multline}

\clearpage
\section{On the energy increase in the GPSL model\label{APPSec:GPSL_EnergyIncrease}}

Here we investigate the energy increase in the GPSL model. Computing $\dot{E}_t$ using Eq.~\eqref{eq:GPSL_MasterEquation_Particles} we get
\begin{equation}
\dot{E}_t = 
-\frac{\lambda}{m_0}M\ev{\hH}_t
+\frac{\lambda}{m_0}\sum_j \frac{1}{2 m_j}\sum_\beta \int \dd[3]{\bx} \Tr{\prt{\sqrt{\hmu_{r_C} (\bx)} U_G^\dg (\bx) \hp_{j,\beta}^2 U_G (\bx) \sqrt{\hmu_{r_C} (\bx)}}\rho_t},
\end{equation}
where $\beta=1,2,3$ denotes the spatial directions.

\subsection{Separation of contributions to the energy increase}

Denoting by $\hq$ the collection of position operators, by $\hq_{j,\beta}$ the position operator of the $j$-th particle in the $\beta$ direction\footnote{Similarly, $\hp_{j,\beta}$ is the momentum operator of the $j$-th particle in the $\beta$ direction.}, and by $F(q)$ a generic function of the position operators, the first technical result we need is that\footnote{The second mathematical identity stems from the Baker-Campbell-Hausdorff formula. One has that $[\hA,\hB]_n:=[\hA,[\hA,\dots[\hA,\hB]\dots]]$ with $n$ nested commutators. Also $[\hA,\hB]_0 = \hB$.}
\begin{equation}
\comm{F(q)}{\hp_{j,\beta}} = i\hbar \partial_{\hq_{j,\beta}}F(q),
\quad 
e^{\hA}\hB e^{-\hA}= \sum_{n=0}^{\infty} \frac{1}{n!}[\hA,\hB]_n
\implies
U_G^\dg (\bx) \hp_{j,\beta} U_G (\bx)
=
\hp_{j,\beta} +\hbar r_j \pdv{\hq_{j,\beta}} f_{r_G} (\hbq_j-\bx).
\end{equation}
From this, we get that
\begin{equation}
U_G^\dg (\bx) \hp_{j,\beta}^2 U_G (\bx)
=
\prtq{U_G^\dg (\bx) \hp_{j,\beta} U_G (\bx)}^2
=
\hp_{j,\beta}^2 
+\hbar r_j \acomm{\hp_{j,\beta}}{\pdv{\hq_{j,\beta}} f_{r_G} (\hbq_j-\bx)}
+ \hbar^2 r_j^2 \prtq{\pdv{\hq_{j,\beta}} f_{r_G} (\hbq_j-\bx)}^2.
\end{equation}
We can rewrite the above by transferring\footnote{When we write $\partial_\beta$ we mean the partial derivative with respect to $\bx$, not $\hbq$.} the partial derivative on the coordinate $\bx$:
\begin{equation}\label{APPeq:GPSL_SandwichExpansion}
U_G^\dg (\bx) \hp_{j,\beta}^2 U_G (\bx)
=
\hp_{j,\beta}^2 
-\hbar r_j \acomm{\hp_{j,\beta}}{\partial_\beta f_{r_G} (\hbq_j-\bx)}
+ \hbar^2 r_j^2 \prtq{\partial_\beta f_{r_G} (\hbq_j-\bx)}^2,
\end{equation}
$\partial_\beta$ derives with respect to $\bx$. 
This allows us to separate the PSL contribution from the exclusively gravitational one. We have
\begin{equation}
\dot{E}_t = \frac{\lambda \hbar^2}{m_0} I_N [g_{r_C};\rho_t] + \mcG[g_{r_C},f_{r_G};\rho_t],
\end{equation}
where
\begin{multline}
\mcG[g_{r_C},f_{r_G};\rho_t]
:=\\
\frac{\lambda}{m_0}\sum_{j,\beta} \frac{1}{2 m_j} \int \dd[3]{\bx} \Tr{\prt{\sqrt{\hmu_{r_C} (\bx)} \prt{-\hbar r_j \acomm{\hp_{j,\beta}}{\partial_\beta f_{r_G} (\hbq_j-\bx)}
			+ \hbar^2 r_j^2 \prtq{\partial_\beta f_{r_G} (\hbq_j-\bx)}^2} \sqrt{\hmu_{r_C} (\bx)}}\rho_t},	
\end{multline}
Indeed, hereafter we will focus on $\mcG[g_{r_C},f_{r_G};\rho_t]$.

We start by analyzing the term $\Tr{\sqrt{\hmu_{r_C} (\bx)} \acomm{\hp_{j,\beta}}{\partial_\beta f_{r_G} (\hbq_j-\bx)} \sqrt{\hmu_{r_C} (\bx)}\rho_t}
$. To do this, we exploit the fact that\footnote{We recall that the derivative only acts on the Dirac delta and not on other objects that may be to the right of the momentum operator.}
\begin{equation}
\mel{q}{\hp_{j,\beta}}{q'}= -i\hbar \prtq{\pdv{q_{j,\beta}}\delta (q-q')},
\end{equation}
and we rewrite the trace as follows:
\begin{multline}
\Tr{\sqrt{\hmu_{r_C} (\bx)} \acomm{\hp_{j,\beta}}{\partial_\beta f_{r_G} (\hbq_j-\bx)} \sqrt{\hmu_{r_C} (\bx)}\hat{\rho}_t}
=\\
=
\int \dd{q} \sqrt{\mu_{r_C} (q;\bx)} \bra{q} \hp_{j,\beta}[\partial_\beta f_{r_G} (\hbq_j-\bx)]\sqrt{\hmu_{r_C} (\bx)}\hat{\rho}_t \ket{q}
+
\int \dd{q} \sqrt{\mu_{r_C} (q;\bx)} [\partial_\beta f_{r_G} (\bq_j-\bx)]\bra{q} \hp_{j,\beta}\sqrt{\hmu_{r_C} (\bx)}\hat{\rho}_t \ket{q}
=\\
=
2\hbar \int \dd{q}\dd{q'} \sqrt{\mu_{r_C} (q;\bx)}\sqrt{\mu_{r_C} (q';\bx)} [\partial_\beta f_{r_G} (\bq_j'-\bx)]\Im[\rho_t(q';q)]\prtq{\pdv{q_{j,\beta}}\delta (q-q')}
=\\
=
-2\hbar \int \dd{q}\dd{q'} \delta (q-q')\pdv{q_{j,\beta}}\prtq{\sqrt{\mu_{r_C} (q;\bx)}\sqrt{\mu_{r_C} (q';\bx)} [\partial_\beta f_{r_G} (\bq_j'-\bx)]\Im[\rho_t(q';q)]}
=\\
=
-2\hbar \int \dd{q}\dd{q'} \delta (q-q') \mu_{r_C} (q';\bx) [\partial_\beta f_{r_G} (\bq_j'-\bx)]\Im[\pdv{q_{j,\beta}}\rho_t(q';q)]
=\\
=
2\hbar \int \dd{q}\dd{q'} \delta (q-q')\mu_{r_C} (q;\bx) [\partial_\beta f_{r_G} (\bq_j-\bx)]\Im[\pdv{q_{j,\beta}}\rho_t(q;q')]
=
2\hbar \int \dd{q} \mu_{r_C} (q;\bx) [\partial_\beta f_{r_G} (\bq_j-\bx)]\mcK_{j,\beta}(q;q),
\end{multline}
where we exploited the distributional property $\partial_{q_{j,\beta}}\delta (q-q')=-\partial_{q'_{j,\beta}}\delta (q-q')$, the term with the derivative of $\sqrt{\mu_{r_C}(q;\bx)}$ is missing because $\Im[\rho_t (q;q)]=0$, and we defined $\mcK_{j,\beta}(q;q'):= \Im[\partial_{q_{j,\beta}}\rho_t(q;q')]$. Regarding the second term, we simply have that
\begin{multline}
\Tr{\sqrt{\hmu_{r_C} (\bx)}\prtq{\partial_\beta f_{r_G} (\hbq_j-\bx)}^2 \sqrt{\hmu_{r_C} (\bx)}\rho_t}
=
\Tr{\prtq{\partial_\beta f_{r_G} (\hbq_j-\bx)}^2 \hmu_{r_C} (\bx)\rho_t}
=\\
=
\int \dd{q} \prtq{\partial_\beta f_{r_G} (\bq_j-\bx)}^2 \mu_{r_C} (q;\bx)\rho_t (q;q).
\end{multline}
So, putting everything together we get
\begin{equation}
\mcG[g_{r_C},f_{r_G};\rho_t]
=
\frac{\lambda}{m_0}\sum_{j,\beta}\frac{\hbar^2 r_j}{2 m_j} \int \dd[3]{\bx}\dd{q} \prtq{\partial_\beta f_{r_G} (\bq_j-\bx)}\mu_{r_C} (q;\bx) \prtg{r_j \prtq{\partial_\beta f_{r_G} (\bq_j-\bx)} \rho_t (q;q) - 2 \mcK_{j,\beta} (q;q)}.
\end{equation}
Expanding the mass density, 
\begin{equation}
\mcG[g_{r_C},f_{r_G};\rho_t]
=
\frac{\lambda}{m_0}\sum_{j,k,\beta}\frac{\hbar^2 r_j m_k}{2 m_j} \int \dd[3]{\bx}\dd{q} \prtq{\partial_\beta f_{r_G} (\bq_j-\bx)}g_{r_C} (\bq_k-\bx) \prtg{r_j \prtq{\partial_\beta f_{r_G} (\bq_j-\bx)} \rho_t (q;q) - 2 \mcK_{j,\beta} (q;q)}.
\end{equation}
We can see that the two terms are characterized by a distinct behavior.
In fact, we can write
\begin{equation}
\mcG[g_{r_C},f_{r_G};\rho_t] = 
\mcG_{0} [g_{r_C},f_{r_G};\rho_t]
+
\mcG_{\lambda} [g_{r_C},f_{r_G};\rho_t],
\end{equation}
where (substituting the explicit form of $r_j$)\footnote{We remark that $\nabla$ applies to the coordinate $\bx$.}
\begin{equation}
\mcG_{0} [g_{r_C},f_{r_G};\rho_t] = -\hbar G \sum_{j,k} m_k \int \dd{q} \dd[3]{\bx} g_{r_C} (\bq_k-\bx) \prtq{\nabla f_{r_G} (\bq_j - \bx)}\cdot \vec{\mcK}_j (q;q),
\end{equation}
with $\vec{\mcK}_j (q;q) = \prt{\mcK_{j,1} (q;q),\mcK_{j,2} (q;q),\mcK_{j,3} (q;q)}$, and
\begin{equation}
\mcG_{\lambda} [g_{r_C},f_{r_G};\rho_t]
=
\frac{1}{2}\frac{G^2 m_0}{\lambda}\sum_{j,k}m_j m_k\int \dd{q} \dd[3]{\bx} g_{r_C} (\bq_k-\bx) \abs{\nabla f_{r_G} (\bq_j-\bx)}^2 \rho_t (q;q).
\end{equation}
So, $\mcG_{0} [g_{r_C},f_{r_G};\rho_t]$ is independent of the collapse rate $\lambda$ while $\mcG_{\lambda} [g_{r_C},f_{r_G};\rho_t]$ goes like $\lambda^{-1}$, exactly as the gravitational decoherence obtained at first order in perturbation in the example analyzed in Ref.~\cite{Piccione2025NewtonianPSL}.

A useful trick to make quick estimations consists of assuming that the state $\rho_t$ is a pure product state with the following form:
\begin{equation}
\rho_t (q;q') = \prod_{j=1}^N e^{-i \bk_j\cdot (\bq_j-\bq'_j)} \phi_j (\bq_j)\phi_j (\bq'_j), 
\quad
\Im[\phi_j (\bq_j)]=0
\implies
\vec{\mcK}_j (q;q) = 
- \bk_j \prod_{j=1}^N \abs{\phi_j (\bq_j)}^2
=
- \bk_j \rho_t (q;q).
\end{equation}
Also, one can assume that the states are sharply localized at great distances so that $g_{r_C} (\bq_k - \bx) \rightarrow \delta (\bq_k - \bx)$, $\nabla f_{r_G} (\bq_j -\bx) \rightarrow -(\bq_j -\bx)/\abs{\bq_j -\bx}^3$, and $\rho_t (q;q) = \prod_l \delta (\bq_l-\bx_l)$, where $\bx_l$ is the position of the $l$-th particle. In this case, one obtains
\begin{equation}\label{APPeq:GPSL_NewtonianForces}
\mcG_{0} [g_{r_C},f_{r_G};\rho_t]
\simeq
\sum_{j} \prtq{\sum_k G m_k m_j \frac{\bx_k-\bx_j}{\abs{\bx_k-\bx_j}^3}}\cdot \bv_j,
\qquad
\bv_j := \frac{\hbar \bk_j}{m_j},
\end{equation}
which is the Newtonian work flux of the gravitational field on the point masses. More precisely, for each $j$ the term in parentheses is the total Newtonian force exerted by the other particles on the $j$-th particle, and $\bv_j$ is its velocity.
Of course, in the above equation, the terms with $j=k$ are intended to be vanishing.

\subsection{Gravitational potential energy}
Eq.~\eqref{APPeq:GPSL_NewtonianForces} makes us suspect of a connection with the rate of change of potential energy. Indeed, classically, one has
\begin{equations}
V(q)&=-\frac{G}{2}\sum_{j \neq k} m_j m_k \frac{1}{\abs{\bx_j-\bx_k}},
\\
\dv{t} V(q)
&=
\frac{G}{2}\sum_{j,k} m_j m_k \frac{(\bx_j-\bx_k)}{\abs{\bx_j-\bx_k}^3}\cdot\prt{\bv_j-\bv_k}
=
-\sum_j \prtq{\sum_k G m_j m_k \frac{(\bx_k-\bx_j)}{\abs{\bx_k-\bx_j}^3}} \bv_j,
\end{equations}
which is, indeed, the opposite of Eq.~\eqref{APPeq:GPSL_NewtonianForces}. Then, let us introduce the operator that one would naturally associate to the gravitational potential energy
\begin{equation}
\hV_{r_C,r_G} 
:= \int \dd[3]{\bx}\dd[3]{\by} \prtq{\frac{-G}{\abs{\bx-\by}}} \frac{\hmu_{r_C} (\bx)\hmu_{r_G}(\by)}{2}
= -\frac{G}{2}\sum_j m_j\int \dd[3]{\bx} \hmu_{r_C} (\bx) f_{r_G} (\hbq_j-\bx)
= -\frac{G}{2}\int \dd[3]{\bx} \hmu_{r_C} (\bx) \hat{F}_{r_G} (\bx),	
\end{equation}
where $\hmu_{r_G}(\by) = \sum_j m_j g_{r_G} (\hbq_j-\by)$ and introduced $\hat{F}_{r_G} (\bx):= \sum_j m_j f_{r_G} (\hbq_j-\bx)$.
Considering, as we did before, a generic system Hamiltonian, we have
\begin{equation}\label{APPeq:GravitationalPotentialEnergyEquivalence}
\hH = \sum_j \frac{\hbp^2_j}{2 m_j} + V(\hat{q})
\implies
\dv{t}\ev{\hV_{r_C,r_G}}_t
=
\frac{i}{\hbar}\ev{\comm{\hH}{\hV_{r_C,r_G}}}_t
=
-\mcG_{0} [g_{r_C},f_{r_G};\rho_t].
\end{equation}
To show this, we first rewrite $\mcG_{0} [g_{r_C},f_{r_G};\rho_t]$ as
\begin{equation}
\mcG_{0} [g_{r_C},f_{r_G};\rho_t] 
= -\hbar G \int \dd{q} \dd[3]{\bx} \mu_{r_C} (q ; \bx) \sum_j \prtq{\nabla f_{r_G} (\bq_j - \bx)}\cdot \vec{\mcK}_j (q;q).
\end{equation}
Then, we compute
\begin{equation}
\frac{i}{\hbar}\comm{\hH}{\hV_{r_C,r_G}}
=
\frac{i}{\hbar}\sum_j \comm{\frac{\hbp_j^2}{2 m_j}}{\hV_{r_C,r_G}}
=
\frac{1}{2}\sum_j \frac{1}{m_j} \acomm{\nabla_{\hbq_j}\hV_{r_C,r_G}}{ \cdot \hbp_j},
\end{equation}
where we used the identity $[\hbp_{j}^2,F(\hq)]=-i\hbar\acomm{\nabla_{\hbq_{j}}F(\hq)}{\cdot \hbp_{j}} = \sum_\beta -i\hbar\acomm{\partial_{\hq_{j,\beta}}F(\hq)}{\hp_{j,\beta}}$. 
We also want to make use of the following identity
\begin{equation}
\Tr\prtq{\acomm{\vec{F}(\hq)}{\cdot \hbp_j}\rho_t}
=2 \hbar \int \dd{q} \vec{F}(q) \cdot \vec{\mcK}_j (q;q),
\qquad
\vec{\mcK}_j (q;q') = \nabla_{\bq_j} \Im[\rho_t (q;q')],
\end{equation}
which we now prove using $\mel{q}{\hp_{j,\beta}}{q'}= -i\hbar \prtq{\partial_{q_{j,\beta}}\delta (q-q')}$ component-wise:
\begin{multline}
\Tr\prtq{\acomm{F(\hq)}{\hp_{j,\beta}}\rho_t}
=
\Tr\prtq{F(\hq)\hp_{j,\beta}\rho_t + \hp_{j,\beta}F(\hq)\rho_t}
=
\Tr\prtq{\hp_{j,\beta}\rho_t F(\hq) + F(\hq)\rho_t\hp_{j,\beta}}
=\\
=
\int \dd{q}\dd{q'}
\prtq{\mel{q}{\hp_{j,\beta}}{q'}\rho_t (q';q)F(q) + F(q)\rho_t (q;q')\mel{q'}{\hp_{j,\beta}}{q}}
=\\
=
\int \dd{q}\dd{q'} F(q)
\prtq{\mel{q}{\hp_{j,\beta}}{q'}\rho_t^* (q;q')-\rho_t (q;q')\mel{q}{\hp_{j,\beta}}{q'}}
=\\
=
i\hbar \int \dd{q}\dd{q'} F(q) \prtq{\partial_{q_{j,\beta}}\delta (q-q')} \prtq{\rho_t (q;q') -  \rho_t^* (q;q')}
=\\
=
-i\hbar \int \dd{q}\dd{q'} F(q) \delta (q-q') \partial_{q_{j,\beta}}\prtq{\rho_t (q;q') -  \rho_t^* (q;q')}
=
2 \hbar \int \dd{q} F(q) \mcK_{j,\beta} (q;q).
\end{multline}
Then, we have
\begin{multline}
\Tr{\frac{i}{\hbar}\comm{\hH}{\hV_{r_C,r_G}}\rho_t}
=
\sum_j \frac{1}{2 m_j} \Tr \prtq{\acomm{\nabla_{\hbq_j} \hV_{r_C,r_G}}{\cdot \hbp_j}\rho_t}
=
\sum_j \frac{\hbar}{m_j} \int \dd{q} \prtq{\nabla_{\hbq_j} \hV_{r_C,r_G} (q)} \cdot \vec{\mcK}_j (q;q)
=\\
=
-\frac{\hbar G}{2}\sum_j \frac{1}{m_j} 
\int \dd[3]{\bx}\dd{q} \prtg{\hF_{r_G} (q;\bx) \prtq{\nabla_{\hbq_j} \hmu_{r_C} (q;\bx)} \cdot \vec{\mcK}_j (q;q)
	+
	\hmu_{r_C} (q;\bx)\prtq{\nabla_{\hbq_j} \hF_{r_G} (q;\bx)} \cdot \vec{\mcK}_j (q;q)}.
\end{multline}
We see that the second addendum in the second line is half of $-\mcG_0$ after expanding $\hF_{r_G}$ and changing $\nabla_{\bq_j}$ with $-\nabla_\bx$. It rests to show that the first addendum integrates to the same quantity. To do this, let us focus on the integral:
\begin{multline}
\sum_{j,\beta} \frac{1}{m_j} \int \dd[3]{\bx}\prtq{\partial_{q_{j,\beta}} \prtq{\mu_{r_C} (q;\bx) F_{r_G} (q;\bx)}}
=\\
=
\sum_{j,\beta} \frac{1}{m_j} \int \dd[3]{\bx}\prtg{
	\prtq{\partial_{q_{j,\beta}} \mu_{r_C}(q;\bx)} F_{r_G} (q;\bx)
	+
	\prtq{\partial_{q_{j,\beta}}F_{r_G} (q;\bx)} \mu_{r_C}(q;\bx)
}
=\\
=
-\sum_{j,\beta} \int \dd[3]{\bx}\prtg{
	\prtq{\partial_\beta g_{r_C}(\bq_j-\bx)} F_{r_G} (q;\bx)
	+
	\prtq{\partial_\beta f_{r_G} (\bq_j-\bx)} \mu_{r_C}(q;\bx)
}
=\\
=
-\sum_{j,\beta} \int \dd[3]{\bx}\prtg{
	\prtq{\partial_\beta f_{r_G} (\bq_j-\bx)} \mu_{r_C}(q;\bx)
	-
	\prtq{\partial_\beta F_{r_G} (q;\bx)}g_{r_C}(\bq_j-\bx) 
}
=\\
=
-\sum_{j,k,\beta} m_k \int \dd[3]{\bx}\prtg{
	\prtq{\partial_\beta f_{r_G} (\bq_j-\bx)} g_{r_C}(\bq_k-\bx)
	-
	\prtq{\partial_\beta f_{r_G} (\bq_k-\bx)}g_{r_C}(\bq_j-\bx) 
}.
\end{multline}
Defining $\bd_{j,k}:=\bq_j-\bq_k$, using $\bx=\bq_k+\by$, and the spherical symmetry of the distributions, we get\footnote{We recall that $\partial_\beta$ is the partial derivative with respect to coordinates. So, $\partial_\beta f(\bd-\by)=-\partial_{\bd,\beta}f(\bd-\by)$.}
\begin{multline}
\int \dd[3]{\bx}\prtq{\partial_\beta f_{r_G} (\bq_j-\bx)} g_{r_C}(\bq_k-\bx)
=
\int \dd[3]{\by}\prtq{\partial_\beta f_{r_G} (\bd_{j,k}-\by)} g_{r_C}(\by)
=\\
=
-\partial_{\bd_{j,k},\beta}\int \dd[3]{\by}f_{r_G} (\bd_{j,k}-\by) g_{r_C}(\by)
=
\partial_{\bd_{j,k},\beta}\int \dd[3]{\by}f_{r_G} (\by+\bd_{j,k}) g_{r_C}(\by).
\end{multline}
Similarly, using $\bx=\bq_j+\by$, we get
\begin{equation} 
-\int \dd[3]{\bx}\prtq{\partial_\beta f_{r_G} (\bq_k-\bx)}g_{r_C}(\bq_j-\bx)
=
\int \dd[3]{\by}\prtq{\partial_\beta f_{r_G} (\by+\bd_{j,k})}g_{r_C}(\by)
=
\partial_{\bd_{j,k},\beta}
\int \dd[3]{\by} f_{r_G} (\by+\bd_{j,k})g_{r_C}(\by).
\end{equation}
So, the two integrands are equal and this finishes the proof of  Eq.~\eqref{APPeq:GravitationalPotentialEnergyEquivalence}.

\clearpage
\section{The isolated particle optimization problem for the GPSL model\label{APPSec:GPSL_Optimization}} 

Here, we want to minimize the quantity\footnote{We recall that $g_{r_C}$ and $g_{r_G}$ are not the same distribution with a different scale.}
\begin{equation}
I_0^{(G)}[g_{r_C},g_{r_G}] := \frac{1}{2}\int \dd[3]{\bx} g_{r_C} (\bx) \abs{\nabla f_{r_G} (\bx)}^2,
\qquad
f_{r_G} (\bx)
=\int \dd[3]{\by} \frac{g_{r_G} (\by)}{\abs{\bx-\by}}.
\end{equation}
where $g_{r_C} (\bx)$ is an arbitrary radial distribution and $g_{r_G} (\bx)$ is assumed to be radial and to satisfy the following constraints:
\begin{equation}
\int \dd[3]{\bx} g_{r_G} (\bx)  = 1,
\quad
\int \dd[3]{\bx} \bx^2 g_{r_G} (\bx)  = 3 r_G^2,
\quad
g_{r_G} (\bx) \geq 0.
\end{equation}

First, we move to spherical coordinates. The constraints now read
\begin{equation}\label{APPeq:GPSL_Constraints}
4\pi \int_0^{\infty} \dd{r} r^2 g_{r_G} (r) = 1,
\quad
4\pi \int_0^{\infty} \dd{r} r^4 g_{r_G} (r)= 3 r_G^2,
\quad
g_{r_G} (r) \geq 0.
\end{equation}
We also have that
\begin{equation}
f_{r_G} (r) 
= 2\pi \int_0^{\infty} \dd{\mu} \int_0^{\pi} \dd{\theta} \frac{\sin(\theta)\mu^2 g_{r_G} (\mu)}{\sqrt{r^2 + \mu^2 - 2 r \mu \cos(\theta)}}
= \frac{4\pi}{r} \int_0^{r} \dd{\mu} \mu^2 g_{r_G} (\mu)
+ 4\pi \int_{r}^{\infty} \dd{\mu} \mu g_{r_G} (\mu)
\end{equation}
and
\begin{equation}
f_{r_G} (\infty) = 0,
\qquad
f_{r_G} (0) = 4\pi \int_0^{\infty} \mu g_{r_G} (\mu) \dd{\mu},
\qquad
f'_{r_G} (r) = -\frac{4\pi}{r^2}\int_0^{r} \mu^2 g_{r_G} (\mu) \dd{\mu}.
\end{equation}
Then, $I_0^{(G)}$ now reads
\begin{equation}
I_0^{(G)}[g_{r_C},g_{r_G}]
=
2 \pi \int_0^\infty \dd{r} r^2 g_{r_C} (r) \prtq{f'_{r_G} (r)}^2
=
2 \pi \int_0^{\infty} \dd{r} \frac{g_{r_C} (r)}{r^2} Q^2(r),
\qquad
Q (r):= 4\pi\int_0^{r} \mu^2 g_{r_G} (\mu) \dd{\mu}.
\end{equation}
The constraints on $g_{r_G}$ translate to $Q(r)$ as follows:
\begin{equation}
Q(r) \geq 0,\qquad
Q'(r) \geq 0,\qquad
Q(0)=Q'(0)=0, \qquad
Q(\infty)=1, \qquad
\int_0^{\infty} r^2 Q'(r) \dd{r} = 3 r_G^2.
\end{equation}

Assume that there is $R$ such that $Q(R)=1$. Then, we have that $\int_0^R Q'(r) \dd{r} = 1$, and $\int_0^{R} r^2 Q'(r)\dd{r} = 3 r_G^2$. So, the Lagrangian reads
\begin{equation}
\mcL [Q] = \int_0^R \dd{r}\prtq{A(r) Q^2 (r) + \lambda Q'(r) + \mu r^2 Q'(r)} - \lambda - 3 r_G^2 \mu,
\qquad
A(r):=2\pi \frac{g_{r_C} (r)}{r^2} 
\end{equation}
Then, we get
\begin{equation}
\dv{\varepsilon}\mcL[Q+\varepsilon \phi]\vert_{\varepsilon=0}
=
\int_0^R \dd{r} \prtq{\lambda \phi' + \mu r^2 \phi' + 2 A Q \phi}
=
\int_0^R \dd{r} \prtq{2 A Q -\dv{r}\prtq{\lambda + \mu r^2}}\phi =0,
\end{equation}
where, in the integration by parts, the condition $Q(0)=0$ implies that $\phi(0)=0$ and $Q(R)=1$ implies $\phi(R)=0$. From the above, we get
\begin{equation}
2 A(r) Q(r) = 2 \mu r, 
\implies
Q(r) 
\propto 
\frac{r^3}{g_{r_C} (r)}\Bigg\vert_{r \leq R}
\implies
Q(r)
=
\begin{cases}
\frac{g_{r_C} (R)}{R^3} \frac{r^3}{g_{r_C} (r)},
\quad
&0\leq r \leq R,
\\
1 \quad &r > R,
\end{cases}
\end{equation}
from which follows
\begin{equation}
g_{r_G} (r)
=
\frac{g_{r_C} (R)}{4\pi R^3 r^2}\dv{r}\prt{\frac{r^3}{g_{r_C} (r)}}\Theta(R-r),
\end{equation}
where $R$ can be now determined by the variance condition $4\pi \int_0^R \dd{r} r^4 g_{r_G} (r) =3 r_G^2$. This condition can be written as
\begin{equation}
4\pi \int_0^R \dd{r} r^4 g_{r_G} (r)
=
\int_0^R r^2 Q'(r)
=
\prtq{r^2 Q(r)}^R_0 - 2\int_0^{R} \dd{r} r Q(r)
=
R^2 - 2 \frac{g_{r_C} (R)}{R^3}\int_0^R \frac{r^4}{g_{r_C} (r)}
=
3 r_G^2.
\end{equation}

Specializing the above equations to the case in which $g_{r_C} (\bx)$ is a Gaussian, one gets
\begin{equation}\label{APPeq:OptimalGravitationalFeedbackDistribution}
Q(r) 
= 
\begin{cases}
\mcN r^3 e^{r^2/2 r_C^2},
\quad
&0\leq r \leq R,
\\
1, \quad &r > R,
\end{cases}
\qquad
g_{r_G} (r) = \frac{Q'(r)}{4\pi r^2} = \mcN \frac{r^2+3 r_C^2}{4\pi r_C^2}e^{r^2/2 r_C^2}\Theta(R-r),
\end{equation}
where $\mcN= R^{-3} \exp{-R^2/2 r_C^2}$ is a normalization constant coming from $Q(R)=1$.
The value of $R$ can be found by imposing the variance condition\footnote{For the definition of the DawsonF function, see here \url{https://reference.wolfram.com/language/ref/DawsonF.html}.}:
\begin{equation}
R^2 - 2 \frac{g_{r_C} (R)}{R^3}\int_0^R \frac{r^4}{g_{r_C} (r)}
=
3 r_G^2
\implies
R^2 -2 r_C^2 + 6\frac{r_C^4}{R^2}-6\sqrt{2}\frac{r_C^5}{R^3}\text{DawsonF}\prt{\frac{R}{\sqrt{2}r_C}}=3 r_G^2. 
\end{equation}
Exploiting the equality $2\times \text{DawsonF}(z)=\sqrt{\pi}e^{-z^2}\text{erfi}(z)$, one can rewrite the above equation as
\begin{equation}\label{APPeq:GPSL_DifferentialEquationBoundRadius}
2y-2+\frac{3}{y}-\frac{3\sqrt{\pi}}{2}\frac{e^{-y}}{y^{3/2}}\text{erfi}(\sqrt{y}) = 3 \prt{\frac{r_G}{r_C}}^2,
\qquad
R= r_C \sqrt{2 y}.
\end{equation}
The constant $R$ can then be found by numerically solving the above equation for the smallest positive solution.

\clearpage
\section{Calculation of the heating functionals for isolated particles\label{APPSec:HeatingFunctionalCalculation}}

To make a comparison,  we compute $I_0^{(G)}$ when both distributions $g_{r_C} (\bx)$ and $g_{r_G} (\bx)$ are Gaussian and also when $g_{r_C} (\bx)$ is a Gaussian but $g_{r_G} (\bx)$ is the optimal distribution found in Eq.~\eqref{APPeq:OptimalGravitationalFeedbackDistribution}. Let us start by using the expression
\begin{equation}\label{APPeq:UsefulExpressionGravitationalHeatingFunctional}
I_0^{(G)}[g_{r_C},g_{r_G}]
=
2 \pi \int_0^{\infty} \dd{r} \frac{g_{r_C} (r)}{r^2} Q^2(r),
\qquad
Q (r)= 4\pi\int_0^{r} \mu^2 g_{r_G} (\mu) \dd{\mu}.
\end{equation}
By defining $u:= r/r_C$, we can write
\begin{equation}
I_0^{(G)}[g_{r_C},g_{r_G}]
=
\frac{1}{\sqrt{2\pi} r_C^4}\int_0^{\infty} \dd{u} \frac{e^{-u^2/2}}{u^2} Q^2(r_C u).
\end{equation}
Let us now focus on the case in which both smearing distributions are Gaussian. We have that
\begin{equation}
Q (r) 
= 
\erf \prt{\frac{r}{\sqrt{2} r_G}} - \sqrt{\frac{2}{\pi}}\frac{r}{r_G}e^{-r^2/2 r_G^2},
\qquad
Q^2 (r)
=
\erf^2 \prt{\frac{r}{\sqrt{2} r_G}} 
- 2\sqrt{\frac{2}{\pi}} \frac{r}{r_G} \erf \prt{\frac{r}{\sqrt{2} r_G}}e^{-r^2/2 r_G^2}
+ \frac{2 r^2}{\pi r_G^2}e^{-r^2/r_G^2}.
\end{equation}
Therefore, introducing $\eta:= r_C/r_G$ we have that
\begin{equation}
Q^2 (r_C u)
=
\erf^2 \prt{\frac{u \eta}{\sqrt{2}}} 
- 2\sqrt{\frac{2}{\pi}} u \eta \erf \prt{\frac{u \eta}{\sqrt{2}}}e^{-u^2 \eta^2/2}
+ \frac{2 u^2 \eta^2}{\pi}e^{-u^2\eta^2},
\end{equation}
so that we can write
\begin{equation}
I_0^{(G)}[g_{r_C},g_{r_G}]
=
\frac{1}{2 r_C^4}\prt{F_1 (\eta) -2 F_2 (\eta) + F_3 (\eta)},
\end{equation}
where
\begin{equations}
F_1 (\eta)
&:= \sqrt{\frac{2}{\pi}}\int_0^{\infty} \dd{u} \frac{e^{-u^2/2}}{u^2} \erf^2 \prt{\frac{u \eta}{\sqrt{2}}},
\\
F_2 (\eta)
&:= 
\frac{2 \eta}{\pi}\int_0^{\infty} \dd{u} \frac{e^{-u^2/2}}{u}\erf \prt{\frac{u \eta}{\sqrt{2}}}e^{-u^2 \eta^2/2},
\\
F_3 (\eta) 
&:= 
\frac{4 \eta^2}{\pi\sqrt{2\pi}} \int_0^{\infty} \dd{u} e^{-u^2\prt{\eta^2+\frac{1}{2}}} = \frac{2}{\pi} \frac{\eta^2}{\sqrt{1+2 \eta^2}}.
\end{equations}
To evaluate $F_1 (\eta)$ and $F_2 (\eta)$ we can take the derivative with respect to $\eta$ and use the boundary condition $F_1(0)=F_2(0)=0$. Therefore, we get
\begin{equations}
\dot{F}_1 (\eta)
&=
\frac{4}{\pi} \int_0^{\infty}\dd{u} \frac{\erf(u\eta/\sqrt{2})}{u}\exp{-u^2 \frac{1+\eta^2}{2}}
=
\frac{4}{\pi} \text{arcsinh} \frac{\eta}{\sqrt{1+\eta^2}},
\\
\dot{F}_2 (\eta)
&=
\frac{1}{2}\dot{F}_1 (\eta)
+
\frac{2\eta}{\pi}\int_0^{\infty} \dd{u} \prt{\sqrt{\frac{2}{\pi}}-u \eta e^{-u^2\eta^2/2}\erf(u \eta/\sqrt{2})}\exp{-u^2 \frac{1+2\eta^2}{2}}
=
\frac{1}{2}\dot{F}_1 (\eta)
+
\frac{2}{\pi} \frac{1}{1+\eta^2}\frac{\eta}{\sqrt{1+2\eta^2}}.
\end{equations}
Then, we have that
\begin{equation}
F_1 (\eta)-2F_2 (\eta)
=
\int_0^\eta \dd{s} \prtq{\dot{F}_1 (s) - 2\dot{F}_2 (s)}
=
-\frac{4}{\pi} \int_0^{\eta} \dd{s}\frac{1}{1+s^2}\frac{s}{\sqrt{1+2 s^2}}
=
1-\frac{4}{\pi}\arctan\sqrt{1+2\eta^2}.
\end{equation}
Putting everything together, we get
\begin{equation}
\text{Both smearings are Gaussian:}
\qquad
I_0^{(G)}[g_{r_C},g_{r_G}]
=
\frac{1}{2 r_C^4}\prt{1 +\frac{2}{\pi} \frac{\eta^2}{\sqrt{1+2 \eta^2}}-\frac{4}{\pi}\arctan\sqrt{1+2\eta^2}}.
\end{equation}

When $g_{r_G}$ is, instead, the optimal distribution, we get
\begin{equation}
g_{r_G} (r) = \mcN \frac{r^2+3 r_C^2}{4\pi r_C^2}e^{r^2/2 r_C^2}\Theta (R-r)
\implies
Q(r) = \mcN r^3 e^{r^2/2 r_C^2}\Theta(R-r)
+
\Theta(r-R),
\quad
\mcN =\frac{e^{-R^2/2 r_C^2}}{R^3}.
\end{equation}
Re-introducing $y=R^2/(2 r_C^2)$ and $u=r/r_C$ we get
\begin{equation}
Q^2 (r_c u) =
\begin{cases}
	\frac{e^{-2y}}{(2y)^3}u^6 e^{u^2}, &\qq{when} u<\sqrt{2y},\\
	1, &\qq{when} u>\sqrt{2y}.
\end{cases}
\end{equation}
so that
\begin{equation}
I_0^{(G)}[g_{r_C},g_{r_G}]
=
\frac{1}{r_C^4 \sqrt{2\pi}} \prtq{\frac{e^{-2y}}{(2y)^3} \int_0^{\sqrt{2y}} \dd{u} u^4 e^{u^2/2} + \int_{\sqrt{2y}}^{\infty} \dd{u} \frac{e^{-u^2/2}}{u^2}}.
\end{equation}
The above expression can be analytically solved as follows:
\begin{equation}\label{APPeq:GPSL_OptimalHeatingFunctionalClosedExpression}
I_0^{(G)}[g_{r_C},g_{r_G}]
=
\frac{1}{2 r_C^4}\prtq{\text{erf}\left(\sqrt{y}\right)+\frac{3 e^{-2 y} \text{erfi}\left(\sqrt{y}\right)}{8 y^3}+\frac{e^{-y} \left(-\frac{3}{4 y^{5/2}}+\frac{1}{2 y^{3/2}}+\frac{1}{\sqrt{y}}\right)}{\sqrt{\pi }}-1}.
\end{equation}

\clearpage
\section{Minimization Universality Counter-Example for 2 Particles in GPSL\label{APPSec:GPSL_CounterExample}}

Here we consider two particles of equal mass $m$, in a state $\rho_t$ such that, in position representation, $\rho_t(\bq_1,\bq_2;\bq_1,\bq_2)=\delta(\bq_1)\delta(\bq_2-\bz)$. We will see that the $g_{r_G}$ found in Appendix~\ref{APPSec:GPSL_Optimization} is not optimal for all values of $\bz$.

The gravitational energy rate contribution is given by
\begin{equation}
\mcG_{\lambda} [g_{r_C},g_{r_G};\rho_t]
=
\frac{G^2 m_0 m^2}{\lambda} \prtg{2 I^{(G)}_{0}[g_{r_C},g_{r_G}] + 2 I^{(G)}_{1,2}[g_{r_C},g_{r_G};\rho_t]},
\end{equation}
Here, $g_{r_C} (\bx)$ is a given radial distribution, so we can write both quantities in spherical coordinates. First, we have [see Appendix~\ref{APPSec:GPSL_Optimization}]
\begin{equation}
I_0^{(G)}[g_{r_C},g_{r_G}]
=
2 \pi \int_0^\infty \dd{r} r^2 g_{r_C} (r) \prtq{f'_{r_G} (r)}^2
=
2 \pi \int_0^{\infty} \dd{r} \frac{g_{r_C} (r)}{r^2} Q^2(r)
=
\int_0^{\infty} \dd{r} w_0 (r) Q^2(r),
\quad
w_0 (r):=2\pi \frac{g_{r_C} (r)}{r^2},
\end{equation}
where $Q (r):= 4\pi\int_0^{r} \mu^2 g_{r_G} (\mu) \dd{\mu}$.
Second we have (with $z=\abs{\bz}$ and setting $\bz=z \hat{e}_3$)
\begin{equation}
I^{(G)}_{1,2}[g_{r_C},g_{r_G};\rho_t]
=
\pi \int_0^{\infty} \dd{r} \int_0^{\pi} \dd{\theta} \sin(\theta) g_{r_C} \prt{\sqrt{z^2+r^2 - 2 z r \cos(\theta)}}\frac{Q^2(r)}{r^2}
=
\int_0^{\infty} \dd{r} w_z (r) Q^2(r),
\end{equation}
where
\begin{equation}
w_z (r) = \frac{\pi}{r^2} \int_0^{\pi} \dd{\theta} \sin(\theta) g_{r_C} \prt{\sqrt{z^2+r^2 - 2 z r \cos(\theta)}}.
\end{equation}
Then, the quantity we would like to minimize is
\begin{equation}
I^{(G)}_{0}[g_{r_C},g_{r_G}] + I^{(G)}_{1,2}[g_{r_C},g_{r_G};\rho_t]
=
\int_0^{\infty} \dd{r} \prtq{w_0 (r) + w_z (r)} Q^2(r),
\end{equation}
which, following the same steps as for the minimization of just $I_0^{(G)}$ will give the result
\begin{equation}
Q(r) = 
\begin{cases}
	\frac{\mu r}{w_0 (r) + w_z (r)} \qq{for} &r < R(z),
	\\
	1, \qq{for} &r \geq R(z),
\end{cases}
\end{equation}
where $R(z)$ will be fixed analogously as in Appendix~\ref{APPSec:GPSL_Optimization}. Unless $w_z (r) = w_0 (r)$, which happens only when $z=0$, this gives a different result than the minimization of $I_0^{(G)}$.

%
%

\clearpage
\section{Calculations for the heating rate of a macroscopic body in GPSL\label{APPSec:GPSL_MacroscopicBodyHeatingRate}}

Here, we simplify the calculation of the integral in Eq.~\eqref{eq:GPSL_MacroscopicBodyGravitationalContributionApproximated} of the main text. 

First, we notice that
\begin{equation}
I_{r_G}
=
\int \dd[3]{\by} I^{(G)} (\by)
=
\frac{1}{2}\int \dd[3]{\bz} \dd[3]{\by} g_{r_C} (\by+\bz) \abs{\nabla f_{r_G} (\bz)}^2
=
\frac{1}{2}\int \dd[3]{\bz}\abs{\nabla f_{r_G} (\bz)}^2,
\qq{where}
f_{r_G} (\bz)
=
\int \dd[3]{\by} \frac{g_{r_G} (\by)}{\abs{\bz-\by}}.
\end{equation}
Since $g_{r_G} (\bx)$ has to be a radial function, we can write
\begin{equation}
\nabla f_{r_G} (\bz)
=
-\int \dd[3]{\by} \frac{\bz-\by}{\abs{\bz-\by}^3}g_{r_G} (\by)
\implies
I_{r_G}
=
\frac{1}{2}\int \dd[3]{\bx}\dd[3]{\by}\dd[3]{\bz}
\frac{(\bz-\bx)\cdot (\bz-\by)}{\abs{\bz-\bx}^3\abs{\bz-\by}^3}g_{r_G}(\bx)g_{r_G}(\by).
\end{equation}
The integration over $\bz$ can be done by orienting $\bd=\bx-\by$ on the polar axis in spherical coordinates. One gets (writing $d=\abs{\bd}$)
\begin{equation}
\int \dd[3]{\bz}
\frac{\bz \cdot (\bz+\bd)}{\abs{\bz}^3\abs{\bz+\bd}^3}
=
2 \pi \int_0^{\infty} \dd{r} r^2 \int_0^{\pi} \dd{\theta} \sin(\theta) \frac{r^2+r d \cos(\theta)}{r^3 \prt{r^2 + d^2 + 2 r d \cos(\theta)}^{3/2}}
=
2 \pi \int_0^{\infty} \dd{r} \frac{1-\sign(d-r)}{r^2}
=
\frac{4 \pi}{d},
\end{equation}
so that now we have
\begin{equation}
I_{r_G}
=
2\pi \int \dd[3]{\bx}\dd[3]{\by} \frac{g_{r_G}(\bx)g_{r_G} (\by)}{\abs{\bx-\by}}.
\end{equation}
One can again go to spherical coordinates and fixing two radii $r$ and $r'$ compute the spherical average. Multiplying this by $4\pi$ to account for all possible directions, one obtains
\begin{equation}
I_{r_G}
=
2 (2\pi)^{3} \int_0^{\infty} \dd{r}\dd{r'} r r' \prt{r+r'-\abs{r-r'}}g_{r_G}(r)g_{r_G}(r')
=
4 (2\pi)^{3} \int_0^{\infty} \dd{r}\dd{r'} \min(r,r') r r' g_{r_G}(r)g_{r_G}(r').
\end{equation}
Since the function we integrate is symmetric with respect to $r$ and $r'$, we can rewrite the integral as
\begin{equation}
I_{r_G}
=
2\int_0^{\infty} \dd{r} \int_0^{r} \dd{r'} 4(2\pi)^3 \min(r,r') r r' g_{r_G}(r)g_{r_G}(r')
=
(4\pi)^3 \int_0^{\infty} \dd{r} r g_{r_G} (r) \int_0^{r} \dd{r'} (r')^2 g_{r_G}(r').
\end{equation}
Now, we can define $G_{r_G} (r)= 4\pi \int_0^{r} \dd{r'} (r')^2 g_{r_G}(r')$, which is the cumulative radial distribution of $g_{r_G}$, so that 
\begin{equation}
I_{r_G}
=
4\pi \int_0^{\infty} \dd{r} \frac{G_{r_G} (r) G_{r_G}' (r)}{r}.
\end{equation}
Now, keeping in mind that $G_{r_G}(0)=0$ and $G_{r_G} (\infty)=1$, we have that
\begin{equation}
\dv{r} \prt{\frac{G_{r_G}^2 (r)}{r}}
=
\frac{2 G_{r_G} (r) G_{r_G}' (r)}{r} - \frac{G^2_{r_G} (r)}{r^2}
\implies
I_{r_G}
=
2 \pi \int_0^{\infty} \dd{r} \frac{G^2_{r_G} (r)}{r^2}.
\end{equation}

Finding the distribution $g_{r_G} (r)$ that minimizes the above quantity leads to the same problem already solved in Appendix~\ref{APPSec:GPSL_Optimization} but with $g_{r_C} (\bx) = 1$. Therefore, we get that 
$g_{r_G} (r) = \mcN \Theta(R-r)$ where $\mcN$ is a normalization constant, $R$ determines the support of $g_{r_G} (r)$, and they can be both determined through the normalization and variance constraints:
\begin{equation}
\begin{cases}
4 \pi \int_0^{\infty} \dd{r} r^2 g_{r_G} (r) = 1,\\
4 \pi \int_0^{\infty} \dd{r} r^4 g_{r_G} (r) = 3 r_G^2,
\end{cases} 
\implies
\begin{cases}
R= \sqrt{5} r_G,\\
\mcN = \prt{\frac{4}{3}\pi R^3}^{-1},
\end{cases}
\implies
G_{r_G} (r)
=
\begin{cases}
\prt{r/R}^3 \qquad &\qq{for} r \leq R,
\\
1 \qquad &\qq{for} r > R.
\end{cases}
\end{equation}
In this case, one gets that
\begin{equation}
I_{r_G}
=
2 \pi \int_0^{\infty} \dd{r} \frac{G^2_{r_G} (r)}{r^2}
=
2 \pi \prtq{\int_0^R \dd{r} \frac{r^4}{R^6} + \int_R^{\infty} \dd{r} r^{-2}}
=
\frac{12 \pi}{5 \sqrt{5}} r_G^{-1}
\simeq
3.37 \times r_G^{-1}.
\end{equation}
Using instead a Gaussian distribution for $g_{r_G} (\bx)$ one gets that
\begin{equation}
I_{r_G}
=
(4\pi)^3 \int_0^{\infty} \dd{r} r g_{r_G} (r) \int_0^{r} \dd{r'} (r')^2 g_{r_G}(r')
=
2 \sqrt{\pi} r_G^{-1}
\simeq
3.54 \times r_G^{-1}.
\end{equation}

\end{document}